\documentclass[final,review, 3p]{elsarticle}

\usepackage{amssymb}
\usepackage{amsmath}

\usepackage{natbib}
\usepackage{multirow}
\usepackage{subfig}
\usepackage[algoruled,boxed,vlined]{algorithm2e}

\journal{}
\let\today\relax

\begin{document}

\begin{frontmatter}



\title{A multiphase meshfree method for continuum-based modeling of dry and  submerged granular flows}


\author[label1]{E. Jafari-Nodoushan}
\author[label2]{A. Shakibaeinia \corref{cor1}}
\cortext[cor1]{Corresponding author, Email: ahmad.shakibaeinia@polymtl.ca}
\author[label1]{K. Hosseini}
\address[label1]{Dep. Civil Engineering, Semnan University, Semnan, Iran}
\address[label2]{Dep. Civil, Geological and Mining Engineering, Polytechnique Montreal, Montreal, Canada}


\begin{abstract}
We develop and fully characterize a meshfree Lagrangian (particle) model for continuum-based numerical modeling of dry and submerged granular flows. The multiphase system of the granular material and the ambient fluid is treated as a multi--density multi--viscosity system in which the viscous behaviour of the granular phase is predicted using a regularized viscoplastic rheological model with a pressure--dependent yield criterion. The numerical technique is based on the Weakly--Compressible Moving Particle Semi-implicit (WC--MPS) method. The required algorithms for approximation of the effective viscosity, effective pressure, and shear stress divergence are introduced. The capability of the model in dealing with the viscoplasticity is validated for the viscoplastic Poiseuille flow between parallel plates. The model is then applied and fully characterized (in respect to the various rheological and numerical parameters) for dry and submerged granular collapses with different aspect ratios. The numerical results are evaluated in comparison with the available experimental measurement in literature as well as some complementary experimental measurements, performed in this study. The results show the capabilities of the presented model and its potential to deal with a broad range of dry and submerged granular flows. It also revealed the impotent role of the regularization, effective pressure, and shear stress divergence calculation methods on the accuracy of the results.

\end{abstract}


\end{frontmatter}


\section{Introduction}
\label{}
Flow of granular materials plays a critical role in many industrial, geophysical and environmental processes, for example in mining operations, landslides, erosion, debris flow, sediment transport, and planetary surface processes. The flow of this most familiar form of matter remains largely unpredictable due to the complex mechanical behaviors that may resemble those of a solid, liquid or gas in different circumstances. Depending on the velocity, a dry granular flow may have three regimes \cite{midia2004dense, jop2006constitutive} including (1) quasi-static regime with negligible grain inertia, which is often described using soil plasticity models, (2) kinetic (or gaseous) regime which exists when the medium is strongly diluted and the contacts between the grains are infrequent, and (3) dense flow (or liquid) regime (in between first two regimes), where a contact network still exists and grain inertia is dominant. The dense regime is often described by the viscoplastic constitutive relations which are still a matter of debate \cite{jop2006constitutive}, even for a simple dry coarse-grain ($d_g>250 \mu m$) cohesionless granular flow \cite{midia2004dense}. The situation is still more complex when the granular material interacts with an ambient viscous fluid like water.  Predicting these so-called multiphase granular flows is critical to further today's limited understanding of many industrial and geo-envirnmental flows, such as fluvial and coastal sediment transport, and submarine landslide. Fluid inertia, viscosity and pore pressure can significantly affect the mechanical behavior of granular flow in such cases \cite{topin2012collapse, cassar2005submarine}.\\

From a numerical standpoint, the approaches of dealing with the granular flows can be either based on the discrete description or the continuum description of granular material \cite{mangeney2007comparison}. The discrete description deals with the individual granular particles subject to the macroscopic and microscopic forces. In this category of methods, the Discrete Element Method (DEM), whereby individual grains are modeled according to Newton's laws with a contact force model, is becoming widely accepted as an effective method. In combination with the computation fluid dynamics (CFD) techniques, discrete--based has also been applied to the two-phase cases of submerged granular flow \cite{zhang2009simulation, topin2012collapse, LiHuntColonius2012, tomac2013discrete, garoosi2015eulerian}. While, the discrete methods are useful in-depth analysis of granular flows, they are relatively computationally intensive \cite{Rycroft2010}. This limits either the length of a simulation or the number of particles, therefore these methods are not suitable for large scale problems such as predicting natural geophysical hazards. In contrast to the discrete description, the continuum description of granular flow treats the granular material as a body of a complex fluid (rather than individual grains) and solves the conservation of mass, momentum and energy to predict the state of the flow system. A continuum description will enable much larger-scale problems to be tackled, therefore,it have been the method of many of past researches (e.g., \cite{medina20082d, moriguchi2009estimating, moriguchi2009estimating, domnik2013coupling, armanini2014submerged, chauchat2014three, yavari2015robust}) for predicting the general behaviour of granular flows. However, the conventional continuum-based modeling of fluid systems relies on a background grid (mesh) system (e.g., in finite volume and finite element methods), which may cause difficulties when dealing with the post-failure interfacial deformations and fragmentation of granular flows \cite{shakibaeinia2011mesh}.\\

The development of the meshfree particle methods for continuum mechanics such as moving particle semi-implicit (MPS)\cite{koshizuka1996moving} and smoothed particle hydrodynamics (SPH) \cite{gingold1977smoothed} methods have provided the opportunity to deal with large interfacial deformations and fragmentation in continuum simulations. These methods have the particle (Lagrangian) nature of discrete methods, such as DEM, yet they don't have the scalability issue as they deal with the continuum. In particular, MPS, the method of this study, has proved to be successful in many fluid mechanics problems (e.g., in \cite{shakibaeinia2011mps, jabbari2011flow, gambaruto2015computational}). \citet{shakibaeinia2010} proposed a weakly compressible MPS method (WC-MPS) for incompressible flow problems. MPS is also a very versatile method to adapt different constitutive equations. Shakibaeinia and Jin \cite{shakibaeinia2011mesh, shakibaeinia2012lagrangian} developed a WC-MPS model in combination with different viscoplastic constitutive relations for applications, such as sand/slurry jet in water and sediment-water interaction in dam-break, respectively.  Similarly, SPH method has  recently been used for the simulation of granular flow problems, such as soil failure (in combination with a elastic–plastic model) \cite{bui2009numerical},  and sediment transport (in combination with a Bingham plastic model) \cite{khanpour2016, manenti2011sph}. Other particle continuum methods such as the material point method (MPM) \cite{mast2015simulating} and the particle finite element method (PFEM) \cite{zhang2014particle} have also been successful in prediction granular flow behaviour, although they still require a global background mesh system.\\


Despite  the recent advances in MPS (and SPH) modeling of granular flows, most of the researches have been limited to the case-specific problems (mostly either flow driven sediment erosion or dry sediment flow). Furthermore, the rheological properties and the numerical implementation techniques are only partially characterized and described. The motivation of this paper, therefore, is to develop, evaluate and fully characterize a general continuum-based meshfree particle numerical framework, based on WC-MPS \cite{shakibaeinia2010} formulation, for modeling of the dense dry and submerged granular flows. It also aims to further our understanding of dry and submerged granular flows mechanism for particular cases of this study.\\

 The proposed numerical model will be based on  a multiphase model, whereby the system is treated as a multi-viscosity multi-density continuum  and the effective viscosity of the granular phase is modeled using a regularized viscoplastic rheological model with a pressure-dependent yield criterion . We will propose and evaluate novel algorithms for approximation of the effective (apparent) viscosity, effective pressure, and shear stress divergence. Relying on extensive numerical simulations, we will also analyze the scalability of the model and the role of various numerical and rheological parameters (e.g., regularization and  post-failure rheological parameters). The model is first validated for the case of  viscoplastic Poiseuille flow, to evaluate its capability of dealing with the viscoplaticity. It is then applied and evaluated for the cases of dry and submerged granular collapse with various aspect ratios. The granular flow features such as failure mechanism, surface profile, and velocity and viscosity fields will be investigated. This study will provide the first comprehensive and fully-characterized continuum-based particle MPS model for simulation of both dry and submerged granular flows.

\section{Governing equations}
\subsection{Flow equations}
The flow governing equations for a weakly-compressible flow system in the Lagrangian frame, including the mass and momentum conservations, equation of state, and material motion, are given by:

\begin{equation}  \label{eq:NS}
\arraycolsep=1pt\def\arraystretch{1.7}
\left\{ \begin{array}{l}
\frac{1}{\varrho }\frac{{{\rm{D}}\varrho }}{{{\rm{D}}t}} + \nabla  \cdot {\bf{u}} = 0\,\\
\varrho \frac{{{\rm{D}}{\bf{u}}}}{{{\rm{D}}t}} =  \nabla  \cdot {\mathbb{T }} + {\bf{f}}\\
p = f(\varrho )\,\\
\frac{{{\rm{D}}{\bf{r}}}}{{{\rm{D}}t}} = {\bf{u}}\,\,
\end{array} \right.
\end{equation}

where, ${\bf{u}}=(u, v)$ is velocity vector, $t$ is time, ${\varrho}$ is  fluid density, $p$ is thermodynamic pressure, ${\bf{f}}$ represents the body forces (e.g., gravity ${\bf{g}}$), ${\bf{r}}=(x, y)$ is the position vector and ${\mathbb{T}}$ is the total stress tensor. Note that in the Lagrangian frame there is no convective acceleration term in the mass and momentum conservation equations, and the material motion is simply calculated by ${{{\rm{D}}{\bf{r}}} \mathord{\left/{\vphantom {{{\rm{D}}{\bf{r}}} {{\rm{D}}t}}} \right.\kern-\nulldelimiterspace} {{\rm{D}}t}} = {\bf{u}}$. It is possible to approximate the behaviour of an incompressible fluid with that of a weakly compressible fluid  using a stiff equation of state (EOS) that used the density field to calculate the pressure field \cite{shakibaeinia2010}. Considering the multiphase system of granular material and ambient fluid as multi-density multi-viscosity system \cite{shakibaeinia2012mps} the governing equations (\ref{eq:NS}) are valid for both phases without considering extra multiphase forces (neglecting the surface tension).

\subsection{Granular rheology}
\label{}
Using the standard definitions in tensor calculus, for an arbitrary tensor $\mathbb{A}$, the trace is given by $\rm{tr}(\mathbb{A})$, the transpose by $\mathbb{A}^T$, the first and second invariants by $I_A$ and $II_A$, magnitude by $\left\| {{\mathbb{A}}} \right\| = \sqrt {I{I_{{\bf{A}}}}}  = \sqrt {0.5{\mathbb{A}}:{\mathbb{A}}} $, and the deviator by ${\mathbb{A'}} = {\mathbb{A}} - \frac{1}{\mathfrak{D}}{\rm{tr}}\left( {\mathbb{A}} \right) \mathbb{I}$, where $\mathfrak{D}$ is the number of space dimensions and $\mathbb{I}$ is the unit tensor. Cauchy symmetric total stress tensor for a general viscous compressible fluid is given by:
\begin{equation}
{\mathbb{T}} =  - p{\mathbb{I}} + {\boldsymbol{\tau}}
\end{equation}

\begin{equation}  \label{eq:SS0}
{\boldsymbol{\tau}} = 2{\eta}{\mathbb{E}} + \xi \left( {\nabla  \cdot {\bf{u}}} \right){\mathbb{I}}
\end{equation}

where $\boldsymbol{\tau}$ is shear stress tensor (deviatoric part of the stress tensor),
${\mathbb{E}} \equiv 0.5\left( {\nabla {\bf{u}} + {{\left( {\nabla {\bf{u}}} \right)}^T}} \right)$ is strain rate tensor (${\mathbb{E'}}$ is its deviator),
 ${\eta}$ is the effective (or apparent) viscosity  (${\eta} = {\eta}\left( {\left\| {{\mathbb{E'}}} \right\|,\,p'} \right)$ for the granular phase and ${\eta} = {\eta _f}$ for the fluid phase) , $\xi$ is the second coefficient of viscosity representing a combination of all the viscous effects associated with the volumetric-rate-of-strain,
 $p' =  - \frac{1}{\mathfrak{D}}{\rm{tr}}\left( {\mathbb{T}} \right) = p - \left( {\xi  + \frac{2}{\mathfrak{D}}{\eta}} \right){\rm{tr}}\left( {\mathbb{E}} \right)$ is normal stress or mechanical pressure (where, $\xi  + \frac{2}{\mathfrak{D}}{\eta}$ is the bulk viscosity). Note that the model of study is a weakly compressible model in which ${\rm{tr}}\left( {\mathbb{E}} \right)$ is close to zero.\\

The effective viscosity is given by a rheological model, whose expression depends on the material under study. The rheological models of this study are Herschel--Bulkley (H--B) generalized viscoplastic model and Bingham plastic (B--P) model that have been widely used for modeling of granular flows \cite{coussot1998direct, chen2002runout, ancey2009dam, chauchat2010three}. In H-B and B--P models the material behaves as a rigid body for stresses less than a yield stress, ${\tau _y} = {\tau _y}\left( {p'} \right)$, but flows as a viscous fluid when stresses exceed this yield stress. The H-B model also accounts for the post-failure non-linear behavior of the stress tensor. The effective viscosity of  H-B  model is given by:

\begin{equation} \label{eq.H-B}
\arraycolsep=1pt\def\arraystretch{1.7}
{\eta} = \left\{ \begin{array}{l}
\frac{{{\tau _y}}}{{2\left\| {{\mathbb{E'}}} \right\|}} + {\eta _0}{\left( {\left\| {{\mathbb{E'}}} \right\|} \right)^{\beta - 1}}\,\,\,\,\,\,\,\,\left\| {\boldsymbol{\tau}} \right\| > {\tau _y}\\
\infty \,\,\,\,\,\,\,\,\,\,\,\,\,\,\,\,\,\,\,\,\,\,\,\,\,\,\,\,\,\,\,\,\,\,\,\,\,\,\,\,\,\,\,\,\,\,\,\,\,\,\,\,\,\,\left\| {\boldsymbol{\tau}} \right\| \le {\tau _y}
\end{array} \right.
\end{equation}

where $\eta _0$ and $\beta$ are the flow consistency and behavior indices (function of material properties such as the grain size and the density and are typically determined through rheometry measurements). $\beta=1$ (case of B--P model) represents the post--failure (after--yield) linear behavior, and  $\beta<1$ account for post--failure shear thickening behaviors. The ideal form of H--B/ B--P models (Eq. \ref{eq.H-B}) is discontinuous and singular for the un-yielded flow regions, when the shear rate approaches zero. (i.e. $\left\| {{\mathbb{E'}}} \right\| \to 0\,;\,\,\left\| {\bf{\tau }} \right\| \to {\tau _y}$). To avoid this singularity (that can cause the numerical model to crash) and also to avoid determining the yielded ($\left\| {\boldsymbol{\tau}} \right\| > {\tau _y}$) and un-yielded ($\left\| {\boldsymbol{\tau}} \right\| \le {\tau _y}$) regions in the flow, a regularized continuous version of the equation is used. The simplest regularization is the bi-viscous model, in which a constant maximum value of ${\eta _{\max }}$ is considered for the effective viscosity (as in \cite{shakibaeinia2011mesh, manenti2011sph, shakibaeinia2012lagrangian, khanpour2016}):

\begin{equation} \label{eq.bi-viscous}
\arraycolsep=1pt\def\arraystretch{1.7}
{\eta} = \left\{ \begin{array}{l}
\frac{{{\tau _y}}}{{2\left\| {{\mathbb{E'}}} \right\|}} + {\eta _0}{\left( {\left\| {{\mathbb{E'}}} \right\|} \right)^{\beta - 1}}\,\,\,\,\,\,\,\,\left\| {\boldsymbol{\tau}} \right\| > {\tau _y}\\
{\eta _{\max }}\,\,\,\,\,\,\,\,\,\,\,\,\,\,\,\,\,\,\,\,\,\,\,\,\,\,\,\,\,\,\,\,\,\,\,\,\,\,\,\,\,\,\,\,\,\,\,\,\,\,\left\| {\boldsymbol{\tau}} \right\| \le {\tau _y}
\end{array} \right.
\end{equation}

\begin{figure}
  \centering
  \includegraphics[width=8cm]{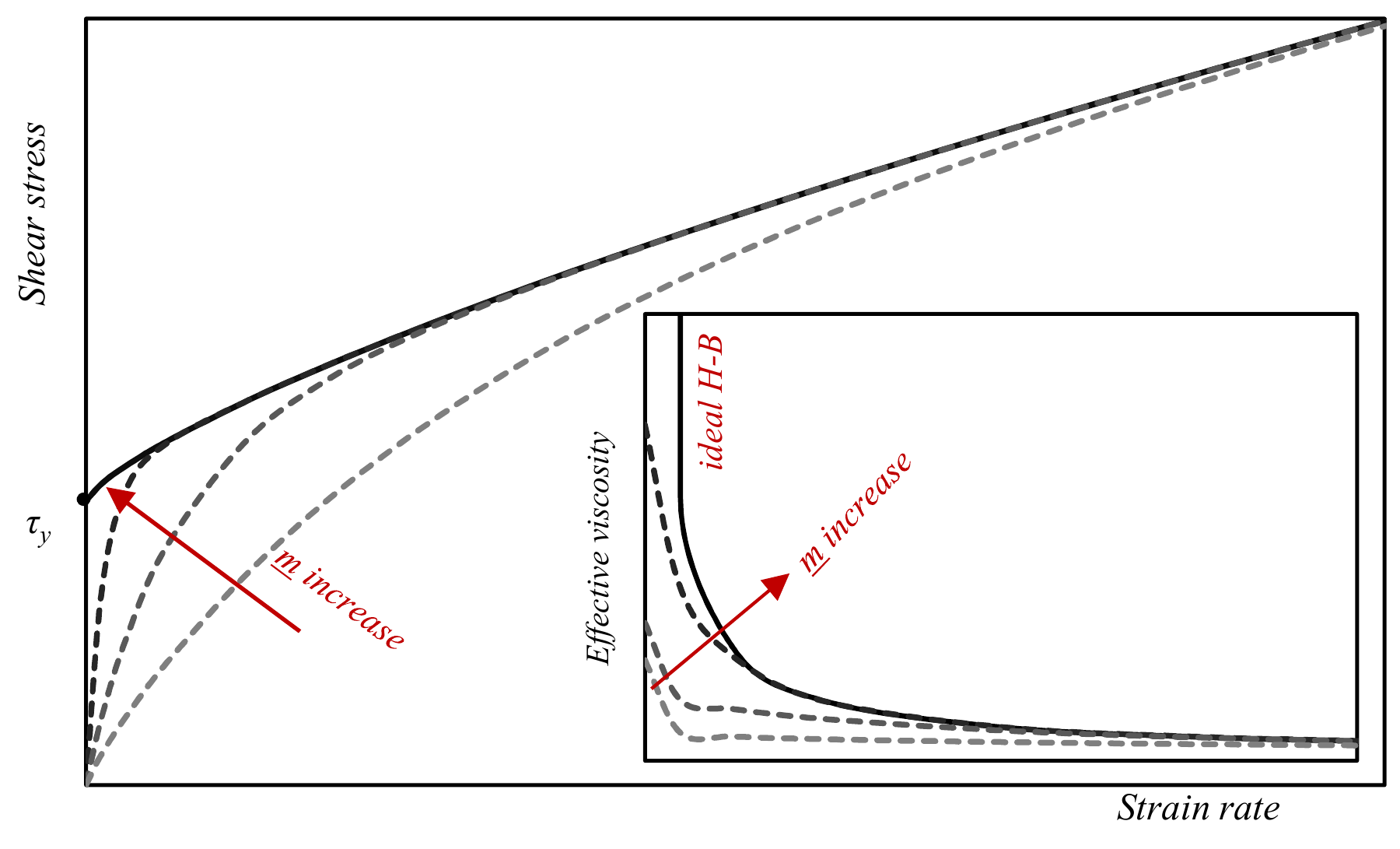}\\
  \caption{Schematic of shear stress and viscosity variation for the H-B model with exponential regularization}\label{fig:reg}
\end{figure}

This paper uses a more popular exponential regularization, proposed by \citet{papanastasiou1987flows}, given by:
\begin{equation} \label{eq.exponential}
{\eta} = \frac{{{\tau _y}\,\left( {1 - \exp ( - m\left\| {{\mathbb{E'}}} \right\|} \right)}}{{2\left\| {{\mathbb{E'}}} \right\|}} + {\eta _0}{\left( {\left\| {{\mathbb{E'}}} \right\|} \right)^{\beta - 1}}
\end{equation}

where parameter $m$ controls the exponential growth of stress, where a larger $m$ value results in a shear stress closer to that of the ideal H-B/B-P model (\ref{fig:reg}). The equation is valid for both yielded and un-yielded regions. For the shear stresses below the yield stress, where $\left\| {{\mathbb{E'}}} \right\|$ approaches zero, the effective viscosity of regularized B--P ($\beta=1$) approaches a maximum viscosity of ${\eta _{\max }} = {\eta _0} + 0.5m{\tau _y}$. Unlike the bi-viscous regularization, this maximum viscosity is a function of yield stress (and normal stress). The yield stress is defined based on the Drucker-Prager  (1952) (or extended von Mises) failure criterion as \cite{chen1996granular}:

\begin{equation}
{\tau _y} = c + {\mu _s}p'
\end{equation}
where c is the cohesiveness coefficient and  ${\mu _s}$ is the threshold coefficient, which corresponds to the material angle of response $\theta _r$. For incompressible cohesionless materials ${\tau _y} = p'\sin {\theta _r}$.

\section{Numerical method}
\label{}
\subsection{MPS fundamentals}
MPS approximation method is based on the local weighted averaging (kernel smoothing) of quantities and vectors (Fig. \ref{fig:kernel}). As a particle method,  MPS represents the continuum  by a set of particles that move in a Lagrangian frame based on the material velocity. Each particle possesses a set of field variables (e.g., mass, velocity, and pressure). Particle $i$ with position vector $\textbf{r}_i$, interacts its neighbor particle $j$ (either in the same phase or different phases) using a weight (kernel) function, ${W\left( {{r_{ij}},{r_e}} \right)}$, where ${r_{ij}} = \left| {{{\bf{r}}_j} - {{\bf{r}}_i}} \right|$ is the distance between particle $i$ and $j$. A third--order polynomial spiky function, proposed by \citet{shakibaeinia2011mesh}, is used in this study. $r_e$ is the radius of the interaction area around each particle. Note that through the course of this paper the word 'particle' refers to a numerical concept, i.e. the MPS computational nodes representing the continuum, on which the governing equation are solved. It should not be mistaken with the physical concept of the material elements (e.g., granular grains or fluid molecules). As a measure of density, a dimensionless parameter, namely particle number density $n$, is defined as:

\begin{figure}
  \centering
  \includegraphics[width=8cm]{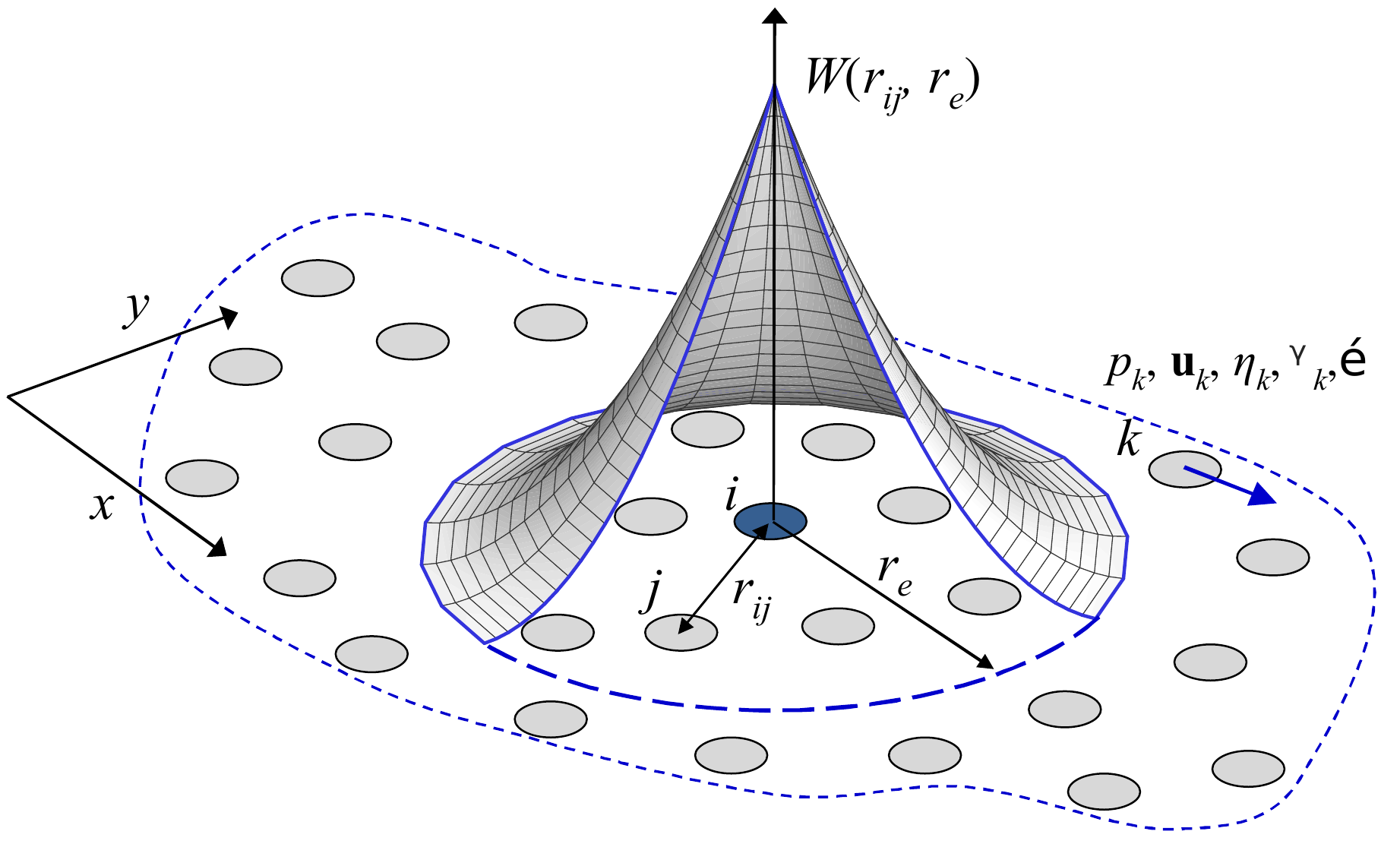}\\
  \caption{MPS kernel approximation}\label{fig:kernel}
\end{figure}

\begin{equation} \label{eq.n}
{\left\langle n \right\rangle _i} = \sum\limits_{j \ne i} {W\left( {{r_{ij}},{r_e}} \right)}
\end{equation}
The operator $\left\langle {} \right\rangle$ is the weight averaging (kernel smoothing or interpolating) operator. Smoothed value of a quantity $f$ can then be calculated by:

\begin{equation}
{\left\langle f \right\rangle _i} = \frac{1}{{n_0 }} \sum\limits_{j \ne i} {{f_j}W\left( {{r_{ij}},{r_e}} \right)}
\end{equation}

where $n_0$ is the averaged value of initial particle number density (a normalization factor). Similarly, the spatial derivatives can be approximated by smoothing the derivatives between the target particles $i$, and each of its neighbors, $j$.  Considering $f$ and $\textbf{F}$ as the arbitrary scalar and vector fields, the gradient of $f$, and divergence and vector-Laplacian of  $\textbf{F}$ are then given by:

\begin{equation}
{\left\langle {\nabla f} \right\rangle _i} = \frac{\mathfrak{D}}{{{n_0}}}\sum\limits_{j \ne i} {\left( {\frac{{{f_i} - {f_i}}}{{{r_{ij}}}}{{\bf{e}}_{ij}}{\mkern 1mu} W\left( {{r_{ij}},{r_e}} \right)} \right)}
\end{equation}

\begin{equation}
{\left\langle {\nabla \cdot{\bf{F}}} \right\rangle _i} = \frac{\mathfrak{D}}{{{n_0}}}\sum\limits_{j \ne i} {\left( {\frac{{{{\bf{F}}_j} - {{\bf{F}}_i}}}{{{r_{ij}}}} \cdot {{\bf{e}}_{ij}}{\mkern 1mu} W\left( {{r_{ij}},{r_e}} \right)} \right)}
\end{equation}

\begin{equation}
{\left\langle {{\nabla ^2}{\bf{F}}} \right\rangle _i} = \frac{{2\mathfrak{D}}}{{\lambda {n_0}}}\sum\limits_{j \ne i} {\left( {\left( {{{\bf{F}}_j} - {{\bf{F}}_i}} \right)W\left( {{r_{ij}},{r_e}} \right)} \right)}
\end{equation}

in which $\mathfrak{D}$ is the number of dimensions, ${{\bf{e}}_{ij}}{{ = {{\bf{r}}_{ij}}} \mathord{\left/
 {\vphantom {{ = {{\bf{r}}_{ij}}} {{r_{ij}}}}} \right.
 \kern-\nulldelimiterspace} {{r_{ij}}}}$ is the unite direction vector and $\lambda$ is a correction parameter by which the increase of numerical variance is equal to that of the analytical solution \cite{koshizuka1996moving} and is defined as:
\begin{equation}
\lambda  = \frac{{\int_V {W(r,{r_e}){r^2}dv} }}{{\int_V {W(r,{r_e})dv} }} \approx \frac{{\sum\limits_{i \ne j} {W\left( {{r_{ij}},{r_e}} \right){r_{ij}}^2} }}{{\sum\limits_{i \ne j} {W\left( {{r_{ij}},{r_e}} \right)} }} = \left\langle {{r_{ij}}^2} \right\rangle
\end{equation}

\subsection{MPS for multiphase granular flow}

\begin{figure}
  \centering
  \includegraphics[width=8cm]{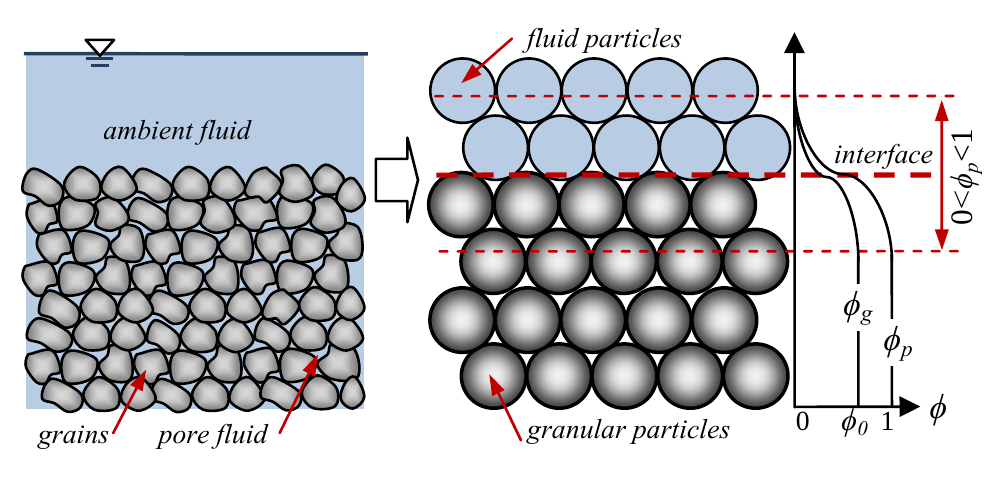}\\
  \caption{Particle representation}\label{fig:particle1}
\end{figure}

Here the physical domain $\Omega  = \left\{ {{\Omega ^f} \cup {\Omega ^g}} \right\}$  ($\Omega ^g $ and $\Omega ^f$ being granular and fluid phases, respectively) is treated as a multi-density multi-viscosity system and a single set of governing equations is solved for the entire flow field.
The solution domain is represented/discretized by the equal-size fluid-type and granular-type particles. As illustrated by Fig. \ref{fig:particle1}, the ambient fluid is represented by the fluid-type particles, and the mixture of granular material and pore/interstitial fluid (trapped inside the porous formed by the grains assembly) is treated as a monophasic isotropic continuum, and is represented by granular-type particles. This is based on the assumption that for a dense granular flow, due to low permeability, the pore/interstitial fluid has negligible mass exchange with the ambient fluid (as it was shown in experimental work of \citet{cassar2005submarine}). The particle size $d_p$ is selected to be equal or greater than the grain size $d_g$ (i.e. the size of the smallest element of the granular continuum).\\

To each of the fluid-type or granular-type particles its own density and viscosity is assigned. The density of the granular continuum is the bulk density of granular matter, and its time/space varying viscosity is determined through the rheological model. A parameter called particles volume fraction (volume concentration), $\phi _p$, describes the volume occupied by granular particles in the unit volume and is define by (for a particle $i$):
\begin{equation} \label{eq.vf}
{\left( {{\phi _p}} \right)_i} =\frac{n_i}{n_0} \frac{{\sum\limits_{j \in \,{\Omega ^g}} {W({r_{ij}},{r_e})} }}{{\sum\limits_{j \ne i} {W({r_{ij}},{r_e})} }}
\end{equation}
The grain volume fraction (the volume occupied by grains in the unit volume) is then given by ${\left( {{\phi _g}} \right)_i} = {\phi _0}{\left( {{\phi _p}} \right)_i}$ (where ${\phi _0}$ is the initial grain volume fraction, assumed to be uniform).The continuous density field for the entire computational domain, is then given by ${\varrho _i}= {\left( {{\phi _g}} \right)_i}{\varrho _g} + \left( {1 - {\left( {{\phi _g}} \right)_i}} \right){\varrho _f}$. This density field, however, smooth the sharp density changes near the interfaces of the granular and fluid phases. Therefore, here we define the density field as:

\begin{equation}
{\varrho_i} = \left\{ \begin{array}{l}
{\varrho _b}= {\phi _0}{\varrho _g} + \left( {1 - {\phi _0}} \right){\varrho _f}\,\,\,\,i \in {\Omega ^g}\\
{\varrho _f}\,\,\,\,\,\,\,\,\,\,\,\,\,\,\,\,\,\,\,\,\,\,\,\,\,\,\,\,\,\,\,\,\,\,\,\,\,\,\,\,\,\,\,\,\,\,\,\,\,\,\,\,\,\,\,\,\,\,\,i \in {\Omega ^f}
\end{array} \right.
\end{equation}

where $\varrho _b$ is the initial bulk density granular continuum, $\varrho _g$ density of grains and $\varrho _f$ is density of ambient fluid. 

\subsubsection{Pressure}
The divergence of the stress tensor equals to the summation of the pressure gradient and the shear stress divergence (Eq. \ref{eq:NS}), written as $\nabla  \cdot {\mathbb{T}} =  - \nabla p + \nabla  \cdot {\boldsymbol{\tau }}$. The MPS approximation of the pressure gradient term $- \nabla p$ for the particle $i$ is given by \cite{koshizuka1996moving}:

\begin{equation} \label{eq.p_grad}
\begin{array}{l}
{\left\langle {\nabla p} \right\rangle _i}\, = \frac{\mathfrak{D}}{{{\varrho _i}{n_0}}}\sum\limits_{i \ne j} {\left[ {\frac{{{p_j} - {{\hat p}_i}}}{{{r_{ij}}}}{{\bf{e}}_{ij}}\,W({r_{ij}},{r_e})} \right]}\\
with\\
{{\hat p}_i} = \mathop {\min }\limits_{} ({p_i},{p_j})\,\,;\,\,j \in \left\{ {W\left( {{r_{ij}},{r_e}} \right) \ne 0} \right\}
\end{array}
\end{equation}

The pressure value is given by the  Tait’s equation of state (EOS) for MPS as \cite{shakibaeinia2010}:
\begin{equation}
p_i^{} = \frac{{{\varrho _i}c_0^2}}{\gamma }\left( {{{\left( {\frac{{{{\left\langle \varrho  \right\rangle }_i}}}{{{\varrho _0}}}} \right)}^\gamma } - 1} \right) = \frac{{{\varrho _i}c_0^2}}{\gamma }\left( {{{\left( {\frac{{{{\left\langle n \right\rangle }_i}}}{{{n_0}}}} \right)}^\gamma } - 1} \right)
\end{equation}

where $c_0$ is the sound speed. The typical value of $\gamma=7$ is used in this relation. To avoid extremely small time steps, a numerical sound speed (smaller than the real sound speed) is selected in a way to keep the compressibility of the fluid less than a certain value. Typically $c_0$ is chosen in order to have Mach number $M < 0.1$, consequently, the density variation is limited to $<0.01$\\

\begin{figure}
  \centering
  \includegraphics[width=8cm]{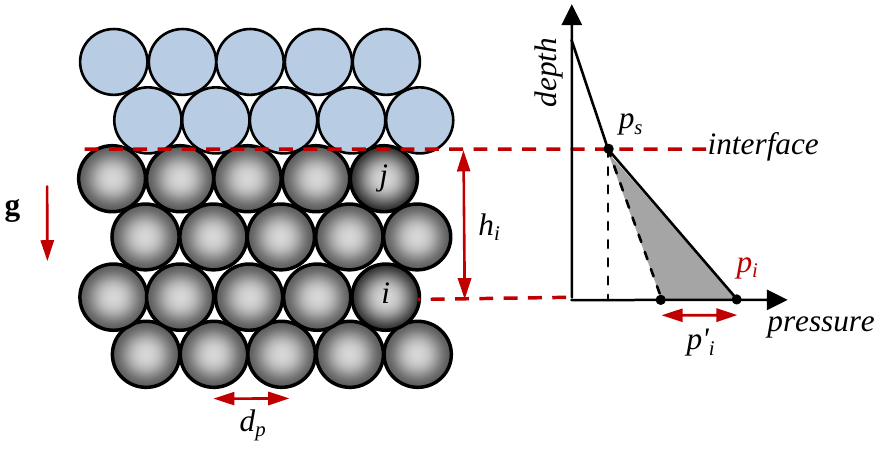}\\
  \caption{Schematic of pressure distribution in submerged granular flow}\label{fig:particle2}
\end{figure}

To calculate the critical shear stress $\tau _y$, the normal stress (mechanical pressure or effective pressure, $p'$) between grains of the granular media is needed. Considering $\xi  =  - \left( {{2 \mathord{\left/{\vphantom {2 D}} \right.\kern-\nulldelimiterspace} \mathfrak{D}}} \right){\eta}$, the normal stress (for dry granular material) will be equal to the thermodynamic pressure, $p' = p$. However, the MPS method is known to be associated with some un-physical pressure fluctuations \cite{shakibaeinia2010}. Such fluctuations (however small) can introduce some un-physical vibrations which can affect the pre- and post-failure behaviours. To avoid this, some recent MPS and SPH studies (e.g., \cite{manenti2011sph, shakibaeinia2011mesh, khanpour2016}), have used a static pressure (for cases with sufficiently small vertical acceleration) that, for the case of dry granular flow, is given by:
\begin{equation}
{{p'}_i} = {\varrho _b}g{h_i}
\end{equation}
where ${h_i}$ is the vertical distance of particle $i$ and the granular surface, given by the locus of granular particles with a particle volume fraction grater than 0.5  (see Fig. \ref{fig:particle2}) defined by:
\begin{equation}
{h_i} = \max \left( {{y_j}} \right) - {y_i}\,\,\,;\,\,j \in \left\{ {\left| {{x_{ij}}} \right| < \delta l,\,and\,\,{\phi _j} > 0.5} \right\}
\end{equation}

However, such static pressure is not applicable in general, where the vertical acceleration may not be negligible. Therefore, in this study, we use the thermodynamic pressure (given by the EOS) but smooth it to minimize the effect of possible pressure fluctuations as:
\begin{equation}
{{p'}_i} = {\left\langle p \right\rangle _i} = \sum\limits_{j \ne i} {{p_j}W({r_{ij}},{r_e})}
\end{equation}

For the submerged granular flow (two-phase flow), the use of the equal-size particles (to represent the granular materials and the ambient fluid) prevents the ambient fluid particles to penetrate the smallest poses between the granular grains. Consequently, the thermodynamic pressure does not account for the fluid pressure in granular proses (pore pressure). Therefore, here the excess pore pressure, $p_{pore}$, must be subtracted (Fig. \ref{fig:particle2}). The effective pressure for the submerged granular flow with static assumption is then given by:

\begin{equation}
\begin{array}{l}
{{p'}_i} = g{h_i}\Delta {\varrho _i}\\
with\\
\Delta {\varrho _i} = {\left( {{\phi _g}} \right)_i}\left( {{\varrho _g} - {\varrho _f}} \right)
\end{array}
 \end{equation}

 For the dynamic condition, the pore pressure, $p_{pore}$, is subtracted from $\left\langle p \right\rangle$ as:


\begin{equation} \label{eq.ep}
\begin{array}{l}
{{p'}_i} =\left\langle p \right\rangle _i  -  \left({p_{pore} }\right)_i= \left( {{{\left\langle p \right\rangle }_i} - {p_s}} \right)\Delta {\varrho _i}/{\varrho _i}\\
with\\
p_s=p_j-\varrho_b g d_p/2
\end{array}
 \end{equation}

where $p_s$ is the pressure at the interface (Fig.\ref{fig:particle2}). Note that for the fully suspended granular particles (where the granular particles have no contacts) ${\left( {{\phi _g}} \right)_i} \to 0$, therefore the normal stress will automatically approach zero, resulting in a zero yield stress.

\subsubsection{Effective viscosity and shear stress}\label{sec.shear_stress}
To calculate the divergence of the shear stress tensor $\nabla  \cdot {\boldsymbol{\tau }}$, the effective viscosity for each particle is required. The effective viscosity of the granular phase particles is given by the constitutive model, while for the fluid phase particles it is a constant value. The effective viscosity for a particle $i$ is then given by:

\begin{equation} \label{eq.ev}
\arraycolsep=1pt\def\arraystretch{1.7}
{\eta_i} = \left\{ \begin{array}{l}
\frac{{\,({\tau _y})_i\left( {1 - \exp ( - m\left\| {{{{\mathbb{E'}}}_i}} \right\|} \right)}}{{2\left\| {{{{\mathbb{E'}}}_i}} \right\|}} + {\eta _0}{\left( {\left\| {{{{\mathbb{E'}}}_i}} \right\|} \right)^{\beta  - 1}}\,\,\,\,i \in {\Omega ^g}\\
{\eta _f}\,\,\,\,\,\,\,\,\,\,\,\,\,\,\,\,\,\,\,\,\,\,\,\,\,\,\,\,\,\,\,\,\,\,\,\,\,\,\,\,\,\,\,\,\,\,\,\,\,\,\,\,\,\,\,\,\,\,\,\,\,\,\,\,\,\,\,\,\,\,\,\,\,i \in {\Omega ^f}
\end{array} \right.
\end{equation}

The MPS approximation of the 2-D strain rate tensor is given by:

\begin{equation} \label{eq.srt}
\arraycolsep=1pt\def\arraystretch{1.5}
\begin{array}{l}
 {\mathbb{E}_i}  = \left[ {\begin{array}{*{20}c}
   {\dot \varepsilon _{xx} } & {\dot \varepsilon _{xy} }  \\
   {\dot \varepsilon _{yx} } & {\dot \varepsilon _{yy} }  \\
\end{array}} \right]_i  \\
  = \frac{1}{{n_0 }}\sum\limits_{j \ne i} {\frac{{W_{ij} }}{{r_{ij}^2 }}\left[ {\begin{array}{*{20}c}
   {2u_{ij} x_{ij} } & {u_{ij} y_{ij}  + v_{ij} x_{ij} }  \\
   {u_{ij} y_{ij}  + v_{ij} x_{ij} } & {2v_{ij} y_{ij} }  \\
\end{array}} \right]}  \\
 \end{array}
 \end{equation}

where $W_{ij}={W\left( {{r_{ij}},{r_e}} \right)}$. By replacing the effective viscosity $\eta_i$ in Eq. \ref{eq:SS0}, the divergence of the shear stress tensor for the particle $i$ can be re-written as:

\begin{equation} \label{eq.SS1}
{\left( {\nabla \cdot{\bf{\tau }}} \right)_i} = \eta_i {\nabla ^2}{{\bf{u}}_i} + \left( {\nabla {{\bf{u}}_i} + {{\left( {\nabla {{\bf{u}}_i}} \right)}^T}} \right)\nabla \eta_i  - \left( {\nabla \cdot{{\bf{u}}_i}} \right)\nabla \eta_i
\end{equation}

in which the MPS approximation of the LHS terms are given by:

 \begin{equation}
\begin{array}{l}
 \left\langle {{\eta _i}{\nabla ^2}{{\bf{u}}_i}} \right\rangle  = \frac{{2\mathfrak{D}}}{{\lambda {n_0}}}\sum\limits_{j \ne i} {\left( {{\eta _i}\left( {{{\bf{u}}_j} - {{\bf{u}}_i}} \right)W({r_{ij}},{r_e})} \right)}  \\
 \left\langle {\nabla  \cdot {\bf{u}}} \right\rangle  = \frac{\mathfrak{D}}{{{n_0}}}\sum\limits_{j \ne i} {\left( {\frac{{\left( {{{\bf{u}}_j} - {{\bf{u}}_i}} \right)}}{{{r_{ij}}}} \cdot {{\bf{e}}_{ij}}\,W({r_{ij}},{r_e})} \right)}  \\
 \left\langle {\nabla {\eta _i}} \right\rangle  = \frac{\mathfrak{D}}{{{n_0}}}\sum\limits_{j \ne i} {\left( {\frac{{\left( {{\eta _j} - {\eta _i}} \right)}}{{{r_{ij}}}}{{\bf{e}}_{ij}}\,W({r_{ij}},{r_e})} \right)}  \\
 \end{array}
\end{equation}

However, when two particles $i$ and $j$ with different effective viscosities of ${\eta _i}$ and ${\eta _j}$ are interacting, the actual effective viscosity of the interaction (friction factor between them) must have a value between ${\eta _i}$ and ${\eta _j}$  (not ${\eta _i}$ ). A harmonic mean value has been recommended by \citet{shakibaeinia2012mps} for this so called interaction viscosity as
${{\mathord{\buildrel{\lower3pt\hbox{$\scriptscriptstyle\frown$}}\over \eta } }_{ij}} = {{\mathord{\buildrel{\lower3pt\hbox{$\scriptscriptstyle\frown$}}\over \eta } }_{ji}} = {{2{\eta _i}{\eta _j}} \mathord{\left/
 {\vphantom {{2{\eta _i}{\eta _j}} {\left( {{\eta _i} + {\eta _j}} \right)}}} \right.
 \kern-\nulldelimiterspace} {\left( {{\eta _i} + {\eta _j}} \right)}}$. Replacing this interaction viscosity in  Eq. \ref{eq.SS1}, the terms including the gradient of the effective viscosity will be eliminated, because $  \left\langle\nabla {{{\mathord{\buildrel{\lower3pt\hbox{$\scriptscriptstyle\frown$}}\over \eta } }_{ij}}}\right\rangle = 0$; therefore, the MPS divergence of the shear stress tensor for particle $i$ is given by:

 \begin{equation}  \label{eq.SS2}
\arraycolsep=1pt\def\arraystretch{1.7}
\begin{array}{l}
{\left\langle {\nabla \cdot{\boldsymbol{\tau }}} \right\rangle _i} = \left\langle {{{{\mathord{\buildrel{\lower3pt\hbox{$\scriptscriptstyle\frown$}}\over \eta } }_{ij}}}{\nabla ^2}{{\bf{u}}_i}} \right\rangle \\
\,\,\,\,\,\,\,\,\,\,\,\,\, = \frac{{4\mathfrak{D}}}{{\lambda {n_0}}}\sum\limits_{j \ne i} {\left( {\frac{{{\eta _i}{\eta _j}}}{{{\eta _i} + {\eta _j}}}\left( {{{\bf{u}}_j} - {{\bf{u}}_i}} \right)W\left( {{r_{ij}},{r_e}} \right)} \right)}
\end{array}
 \end{equation}

 An alternative method, examined in this paper, is to calculate MPS approximation of shear stress tensor first, then calculate its divergence as:
 \begin{equation} \label{eq.SS3}
\left\{ {\begin{array}{*{20}{l}}
{{{\left\langle {\nabla \cdot{\boldsymbol{\tau }}} \right\rangle }_{ix}} = \frac{\mathfrak{D}}{{{n_0}}}\sum\limits_{j \ne i} {\left( {\frac{{{{\left( {{\tau _{xx}}} \right)}_{ij}}{x_{ij}}}}{{r_{ij}^2}}{W_{ij}} + \frac{{{{\left( {{\tau _{xy}}} \right)}_{ij}}{y_{ij}}}}{{r_{ij}^2}}{W_{ij}}{\mkern 1mu} } \right)} }\\
{{{\left\langle {\nabla \cdot{\boldsymbol{\tau }}} \right\rangle }_{iy}} = \frac{\mathfrak{D}}{{{n_0}}}\sum\limits_{j \ne i} {\left( {\frac{{{{\left( {{\tau _{xy}}} \right)}_{ij}}{x_{ij}}}}{{r_{ij}^2}}{W_{ij}} + \frac{{{{\left( {{\tau _{yy}}} \right)}_{ij}}{y_{ij}}}}{{r_{ij}^2}}{W_{ij}}} \right)} }
\end{array}} \right.
\end{equation}

where for the weak compressibility (negligible divergence):
 \begin{equation}
 \arraycolsep=1pt\def\arraystretch{1.5}
\left[ {\begin{array}{*{20}c}
   {\tau _{xx} } & {\tau _{xy} }  \\
   {\tau _{xy} } & {\tau _{yy} }  \\
\end{array}} \right]_i  = \frac{\mathfrak{D}}{{n_0 }}\sum\limits_{j \ne i} {\frac{{\eta _{ij} W_{ij} }}{{r_{ij}^2 }}\left[ {\begin{array}{*{20}c}
   {2u_{ij} x_{ij} } & {v_{ij} x_{ij}  + u_{ij} y_{ij} }  \\
   {v_{ij} x_{ij}  + u_{ij} y_{ij} } & {2v_{ij} y_{ij} }  \\
\end{array}} \right]}
\end{equation}

Nevertheless, this method is not recommended because it involves two times MPS approximation (once for calculation the stress tensor and once for calculation of its divergence), which can smooth the sharp changes in the viscosity and velocity fields. This will be investigated later in this paper.


\subsubsection{Boundary condition}
The free surface boundary condition ($p=0$) is implicitly included into the WC-MPS formulation. For the solid boundaries, some layers of so called ghost particles are placed right outside the computational domain boundaries to compensate the deficiency in the near-boundary particle number density ($n_i$) and to assign the boundary values. Ghost particles are fixed-position and uniformly-distributed with a spacing equal to the initial spacing of the real particles. The ghost particle's pressure and tangential velocity component are extrapolated from the flow domain and the normal velocity is set to zero. the implication of solid boundaries has been discussed in detail in \citet{ shakibaeinia2010, shakibaeinia2011mps}.
\subsubsection{Solution algorithm}

The time integration is based on a  fractional--step method, where each time step is split into two pseudo steps, i.e. the prediction and the correction. The velocity vector for the particle $i$ at the new time step, $t + \delta t$, is achieved by summing the velocity calculated at the prediction and correction steps:
\begin{equation}
{\bf{u}}_i^{t + \delta t} = {\bf{u}}_i^* + {{\bf{u'}}_i}
\end{equation}

where superscripts ``$^{*}$'' and ``$^{'}$'' indicate the predicted and corrected variables, respectively. The viscous force, the body forces and the pressure force, at time $t$, are explicitly used to predict the velocity and pressure force, at time $t + \delta t$. The pressure calculated in a new time step $t + \delta t$ is then used to correct the velocity.

 \begin{equation}
{\bf{u}}_i^* = {\bf{u}}_i^t + \frac{{\delta t}}{{{\varrho _i}}}\left( {{{\bf{f}}_i} + \left( {\nabla  \cdot \tau } \right)_i^t\, - (1 - \alpha )\nabla p_i^t} \right)
\end{equation}
 \begin{equation}
{\bf{u'}} =  - \alpha \frac{{\delta t}}{{{\varrho _i}}}\nabla p_i^{t + \delta t}
\end{equation}

where $\alpha  \in \left[ {0,1} \right]$ is a relaxation factor (typically =0.5). The predicted velocity is used to predict the particle position ${\bf{r}}_i^* = {\bf{u}}_i^*\delta t$ and the particle number density $n_i^*$ (Eq. \ref{eq.n}). The pressure at $t + \delta t$ is then calculated using $n_i^*$, which then is used to calculate the pressure gradient and the corrected velocity $\bf{u'}$. The velocity  ${\bf{u}}_i^{t + \delta t}$ and the particle position ${\bf{r}}_i^{t + \delta t} = \delta t\,{\bf{u}}_i^{t + \delta t}$ are then updated. The time step size $\delta t$ is constrained by CFL (Courant-Friedrichs-Lewy) stability condition, based on both the numerical value of sound speed $c_0$ and the viscous diffusion as:
 \begin{equation}
\delta t = C_r \cdot \min \left( {\frac{{\delta l}}{{{c_0} + \left| {{{\bf{u}}}} \right|_{\max }}},\,\,\frac{{\varrho \,\delta {l^2}}}{{{2\mathfrak{D}}{\eta _{\max }}}}} \right)
\end{equation}

where $C_r\in(0,1]$ is the CFL number. Algorithm 1 summarizes the solution scheme.

\begin{algorithm}
initialization $\textbf{r}_i ,\textbf{u}_i, p_i, \varrho_i$\;
\While{$t + \delta t<T_{final}$}{
Set neighboring table\;
\ForEach{particles i}{
compute volume fraction $\phi$ (Eq. \ref{eq.vf}), strain rate tensor $\mathbb{E}$ (Eq. \ref{eq.srt}), effective pressure $p'$ (Eq. \ref{eq.ep})and yield stress $\tau_y$\;
compute effective viscosity $\eta$ (Eq. \ref{eq.ev})\;
}
\ForEach{particles i}{
compute the viscous (Eq. \ref{eq.SS2}) and pressure forces (Eq. \ref{eq.p_grad})\;
predict velocity $\textbf{u}^*$ and particle position $\textbf{r}^*$\;
}
\ForEach{particles i}{
predict $n^*$ and compute new pressure $p^{t+\delta t}$\;
}
\ForEach{particles i}{
compute new pressure forces and velocity correction\;
update new velocity and particle position\;
apply boundary condition\;
}
$t \leftarrow t + \delta t$
}
\caption{Solution algorithm}
\end{algorithm}

\section{Applications}

\subsection{Viscoplastic Poiseuille flow}
\begin{figure}
  \centering
  \includegraphics[width=8cm]{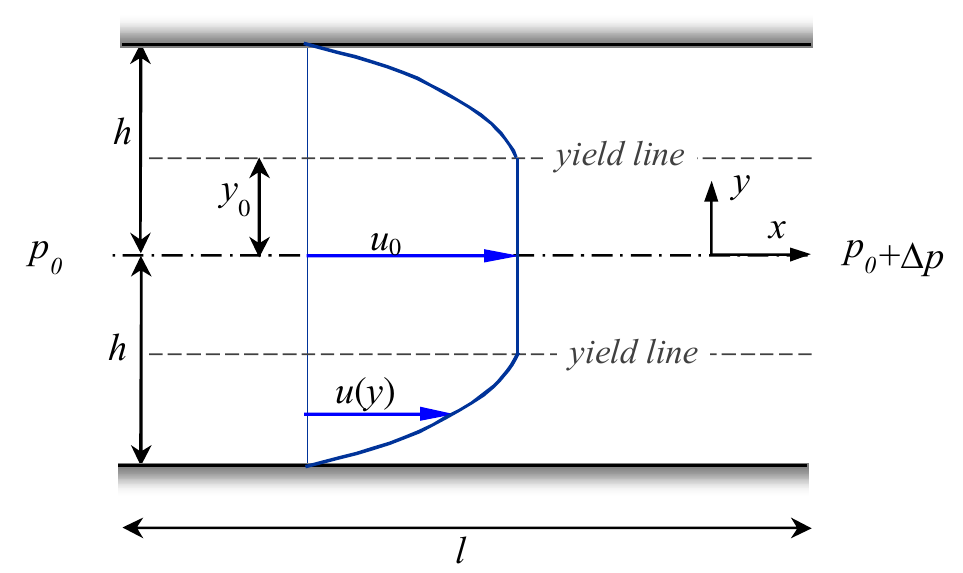}\\
  \caption{Schematic of viscoplastic Poiseuille flow problem}\label{fig:poi_geo}
\end{figure}

Before evaluating the model for the granular flows, the developed model is validated for a benchmark test case, i.e. plane Poiseuille flow of a viscoplastic (B-P) fluid (Fig. \ref{fig:poi_geo}) to evaluate its ability to reproduce the viscoplasticity and yield behaviours. The analytical steady-state velocity field using an ideal B-P rheology is given by:
 \begin{equation}
  \arraycolsep=1pt\def\arraystretch{1.5}
\begin{array}{l}
 u(y) = \left\{ \begin{array}{l}
 u_0 \, = \frac{1}{{4\eta _0 }}f_p\left( {h - y_0 } \right)^2 \,,\,\,\,\,\,\,\,\,\,\,\,\,\,\,\,\,\,\,\,\,\,\,\,\,\,\,\,\,0\, \le y \le y_0  \\
 \frac{1}{{4\eta _0 }}f_p\left( {h^2  - y^2 } \right) - \frac{{\tau _y }}{{2\eta _0 }}\left( {h - y} \right),\,\,\,\,\,\,\,\,y_0  \le y \le h\,\,\,\,\, \\
 \end{array} \right. \\
 v(y) = 0 \\
 \end{array}
\end{equation}
where
 \begin{equation}
f_p =  - \frac{{\partial p}}{{\partial x}} =  - \frac{{\Delta p}}{l};\,\,\,\,\,\,y_0  = \frac{{\tau _y }}{f_p} < h
\end{equation}

\begin{figure}
  \centering
  \includegraphics[width=8cm]{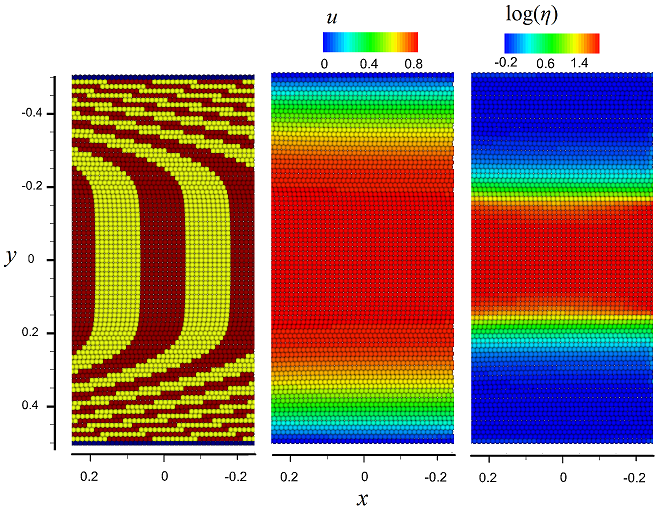}\\
  \caption{Snapshots of particles configuration, velocity and effective viscosity at $t$=2 [unite time] for Poiseuille flow with $\tau_y=1.0$}\label{fig:pois_plot}
\end{figure}

\begin{figure}
  \centering
  \includegraphics[width=8cm]{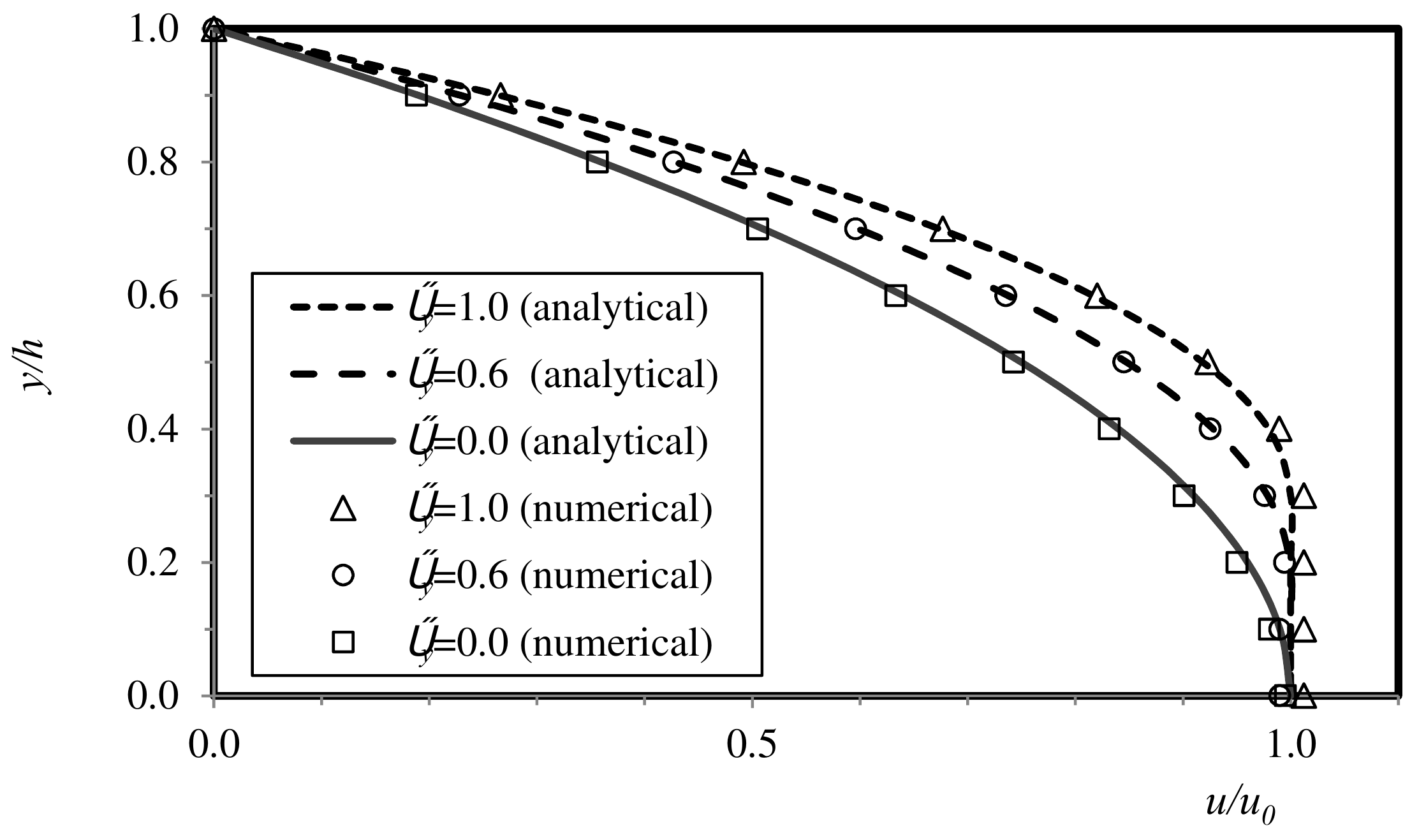}\\
  \caption{Comparison of analytical and numerical streamwise velocity profiles (normalized using analytical centreline velocity $u_0$) of Poiseuille flow with different $\tau_{0}$ values}\label{fig:pois_graph}
\end{figure}

$h$ and $y_0$ are distances of solid boundary and yield line from the centreline, respectively, and $u_0$ is the centreline velocity. Flow occurs only if $f_p>\tau _y h$. The numerical model is applied for a dimensionless domain with $h=0.5$, $\frac{{\Delta p}}{l}=6.667$, $\eta_0=0.25$. Three cases with the constant (pressure-independent) yield stress values of $\tau_y=1.0$, 0.6, and 0.0 are considered. The solution domain is represented by the equal-size particles (with size of $d_p=0.0125$) which have been colored in a way that form some strips (perpendicular to the flow direction) to better show the internal deformations. To expedite the simulations, an initial Newtonian velocity profile is assigned to the particles. An exponentially regularized B-P model with $m=200$ has been used. The model is run until a relatively steady-state condition is achieved (around $t$=2) Fig. \ref{fig:pois_plot} shows a snapshot of the numerical results (for the case with $\tau_y=1.0$) including: the particle configuration, the streamline velocity and the effective viscosity (in logarithmic scale) at $t$=2. As expected, the results show two distinguished, yielded and un-yielded, flow zones. The zone near the side walls (yielded) has a very small effective viscosity and variable velocity, while, the zone near centreline (un-yielded) has a very high effective viscosity and constant velocity. A comparison between the numerical and analytical profiles for various $\tau_y$ values are provided in Fig. \ref{fig:pois_graph}. The agreement between the numerical and analytical results is very good (RSME= 0.008, 0.005 and 0.007 for $\tau_y=1.0$, 0.6, and 0.0, respectively) . A slight over-estimation of velocity near the centreline ($y\leq y_0$) can be due to the insufficiency of the $m$ to avoid deformations (esp. for higher $\tau_y$ values). The results validate the implementation of rheological model and demonstrate the capabilities of the model in reproducing the viscoplasticity and yield behaviours.

\subsection{Granular collapse}
To investigate the role of various rheological and numerical parameters and to evaluate the capabilities of the proposed model, the model is applied investigate the collapse of dry and  submerged (in viscous fluid) granular material, under the gravity force, which mimics the landslide. The setting has been widely used as a benchmark in the past studies (e.g. in \cite{lajeunesse2005, lube2005collapses, balmforth2005, thompson2007granular, cassar2005submarine, rondon2011granular}), because of its simple geometry, the availability of observed/modeled data and the importance in a variety of engineering and geophysics settings.\\

\begin{figure}
  \centering
  \includegraphics[width=8cm]{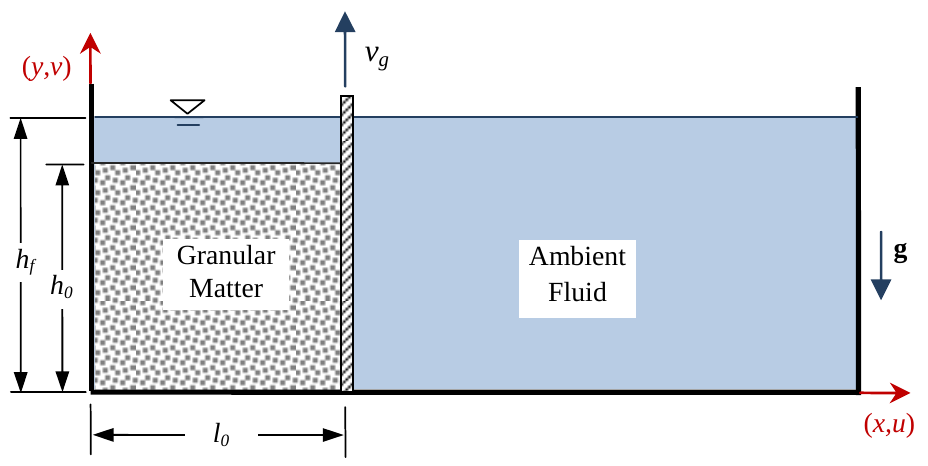}\\
  \caption{Application setup}\label{fig:geo}
\end{figure}

The setup configuration, used in this study, includes a rectangular tank with a horizontal bed (Fig.\ref{fig:geo}), partially filled with granular materials to form a rectangular heap of length $l_0$, height $h_0$, and thickness $w$. The granular column is separated from downstream using a vertical removable gate, creating a reservoir that allows the sudden release of granular mass. The gate can be quickly removed with the vertical speed of $v_g$ to release the granular material. Granular material with the grain density of $\varrho_g$, diameter of $d$, volume fraction of $\phi_0$, and response angle of $\theta_r$ is used. For the  submerged test cases the tank is filled (upstream and downstream of the gate) with a fluid with height of $h_f$, density of $\varrho_f$ and dynamic viscosity of $\eta_f$. The aspect ratio of the initial granular column, $a=h_0/l_0$, is known to be a key factor in the spreading distance of the granular assembly \cite{balmforth2005, lajeunesse2005, thompson2007granular}. Therefore, in this study various aspect ratios are considered. Tables \ref{table:geo} and \ref{table:material} summarize the geometrical and material properties for different settings of this study, respectively.\\

Setups D1 and D2 are based on the experimental settings of \citet{lajeunesse2005} for the dry granular collapse. For the submerged granular collapse, the few available experimental works (e.g., those of \cite{rondon2011granular}) have been done for  small grain size of $d<250\mu m$ and exhibited some cohesion in the earlier stages of collapse, which is against the cohesionless assumption made for the model of this study. Therefore, to validate  the submerged granular model, we have performed a set of experiments on the dense submerged granular flow (setups  S1 and S2), with the geometrical/material properties similar to those of setups D1 and D2, but include an ambient fluid (i.e. water).

\begin{table}[h]\centering
\caption{Geometrical properties}
\label{table:geo}
\begin{tabular}{cccccccc}
\hline
\multirow{2}{*}{Setup} & \multirow{2}{*}{Type}    & $l_0$& $h_0$  &  $a$  & $h_f$ & $v_g$ & \multirow{2}{*}{Validation }
          \\
           &               & \multicolumn{1}{l}{[m]} & \multicolumn{1}{l}{[m]} & \multicolumn{1}{l}{} & \multicolumn{1}{l}{[m]} & \multicolumn{1}{l}{[m/s]} &      \\ \hline
D1    & Dry      & 0.102 & 0.061 & 0.6 & -     & $\sim$ 1.6     & Exp. \cite{lajeunesse2005}    \\
D2    & Dry      & 0.056 & 0.134 & 2.4 & -     & $\sim$ 2.2     & Exp. \cite{lajeunesse2005}    \\
S1    & Submerged & 0.098 & 0.060 & 0.6 & 0.1  & $\sim$ 1.0    & Exp. (present study)                \\
S2    & Submerged & 0.048 & 0.117 & 2.4 & 0.1  & $\sim$ 1.0     & Exp. (present study)             \\\hline
\end{tabular}
\end{table}

\begin{table}[h]\centering
\caption{Material properties}
\label{table:material}
\begin{tabular}{cccccccccc}
\hline
\multirow{2}{*}{Setup} & \multirow{2}{*}{Granular} & $d$                & $\varrho_g$                & $\phi_0$                   & $\theta_r$                   & $\varrho_f$            & $\eta_f$            \\
                       &                           & \multicolumn{1}{l}{[mm]} & \multicolumn{1}{l}{[kg.m\textsuperscript{-3}]} & \multicolumn{1}{l}{} & \multicolumn{1}{l}{[deg.]} & \multicolumn{1}{l}{[kg.m\textsuperscript{-3}]} & \multicolumn{1}{l}{[pa.s]} \\ \hline
D1                     & glass beads               & 1.15                   & 2500                      & $\sim$ 0.63                 & 22                 & -                     & -                     \\
D2                     & glass beads                & 1.15                   & 2500                      & $\sim$ 0.63                  & 22               & -                     & -                     \\
S1                     & glass beads               & 0.80                   & 2500                      & $\sim$ 0.62                  & 22                & 1000                  & 0.001                 \\
S2                     & glass beads                & 0.80                   & 2500                      & $\sim$ 0.62                 & 22                & 1000                  & 0.001                 \\\hline
\end{tabular}
\end{table}

\subsection{Dry granular flow results}
After extensive sensitivity analysis, in respect to the choice of various rheological and numerical parameters and algorithms, the final model (which will be used as a reference model) was archived for simulation  of dry granular collapse.  This reference model has the consistency index of $\eta_0=1.5$ pa.s\textsuperscript{$\beta$}, and the flow behavior index of $\beta=0.4$. A dynamic effective pressure in combination with the Drucker-Prager failure criterion is used to calculate the yield stress. The exponential regularization with $m=50$ and the particle size equal to the grain size ($d_p= d_g$) is used.  A no--slip condition is implemented near solid boundaries. The numerical algorithm is based on Algorithm 1. The characteristic length of $l_0$, (the initial column length), characteristic time of $\surd {h_0/g}$ (i.e. free fall time scale), are used for normalization of the results. In order to better highlight the deformations of the granular assembly, the particles are initially colored differently, to form several vertical layers (similar to \cite{lim2014contact}). Here we first present the simulation result of the reference model, described above, thereafter we investigate the effect of various rheological and numerical parameters and algorithms on the results.

\subsubsection{Flow configuration}

\begin{figure}
  \centering
  \includegraphics[width=\textwidth]{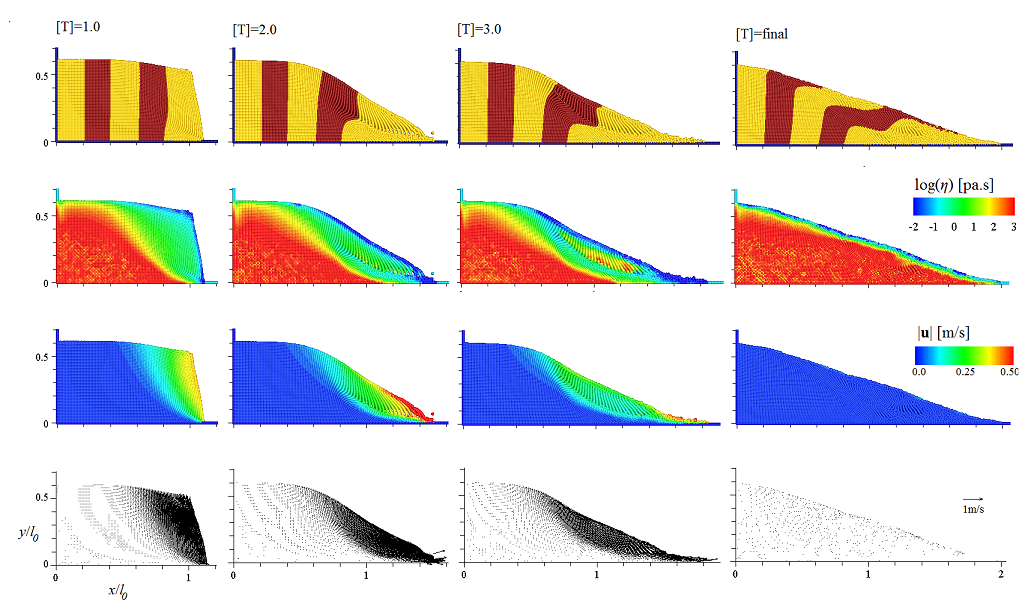}\\
  \caption{Deposit configuration, viscous field and velocity field and vectors for dry granular collapse with $a=0.6$  (test D1)}\label{fig:results_configd1}
  \label{fig:result_all_d1}%
\end{figure}

\begin{figure}
  \centering
  \includegraphics[width=\textwidth]{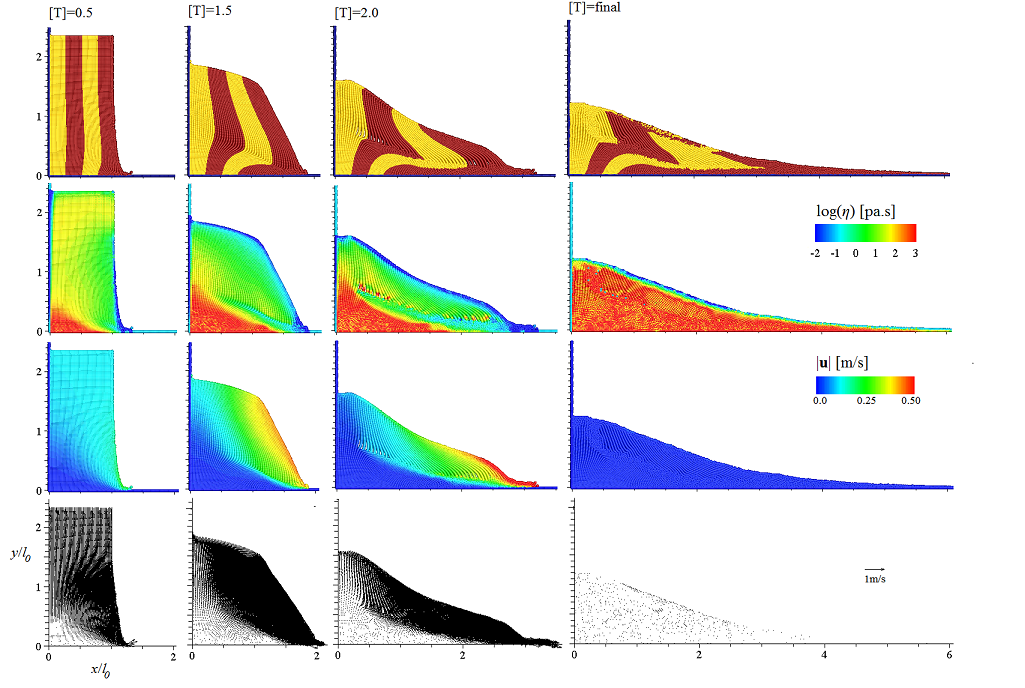}\\
  \caption{Deposit configuration, viscous field and velocity field and vectors for dry granular collapse with $a=2.4$  (test D2)}\label{fig:resultsd_configd2}
  \label{fig:result_all_d2}%
\end{figure}

\begin{figure}%
\centering
\subfloat[][]{\includegraphics[width=6cm]{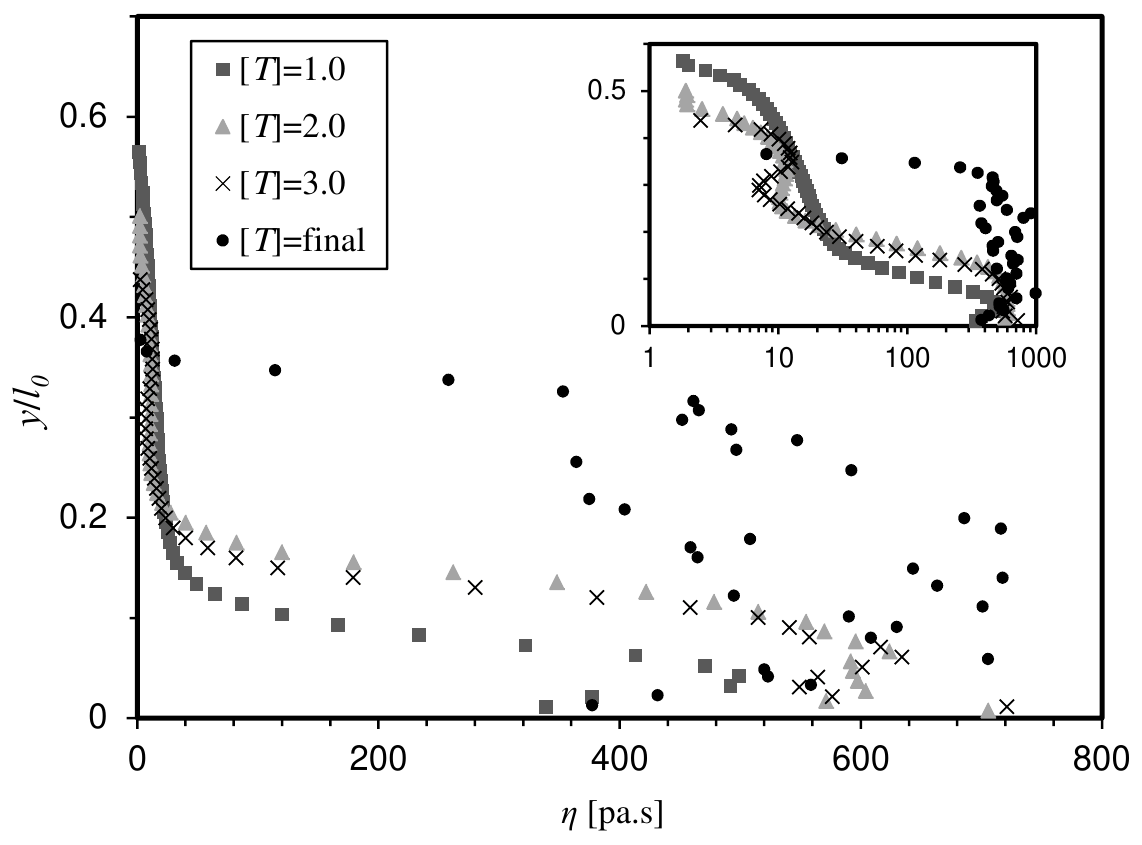}}
\subfloat[][]{\includegraphics[width=6cm]{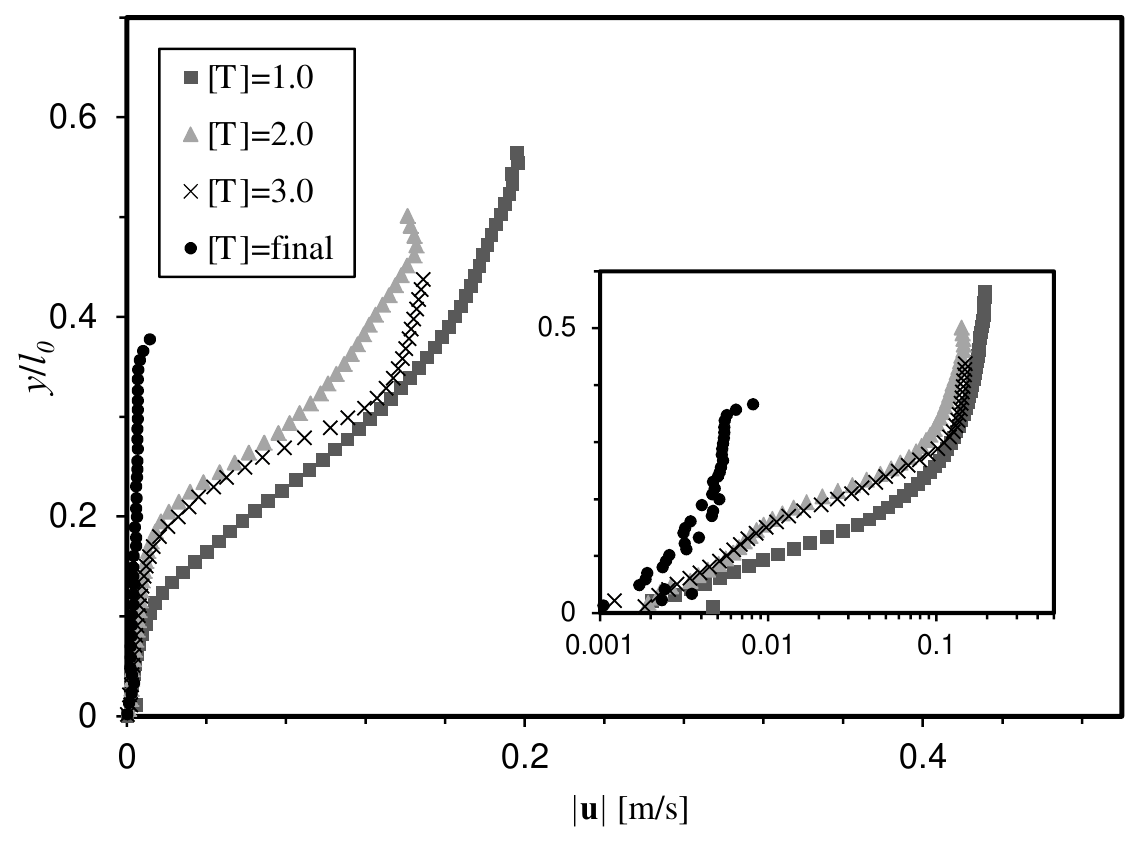}}%
\qquad
\subfloat[][]{\includegraphics[width=6cm]{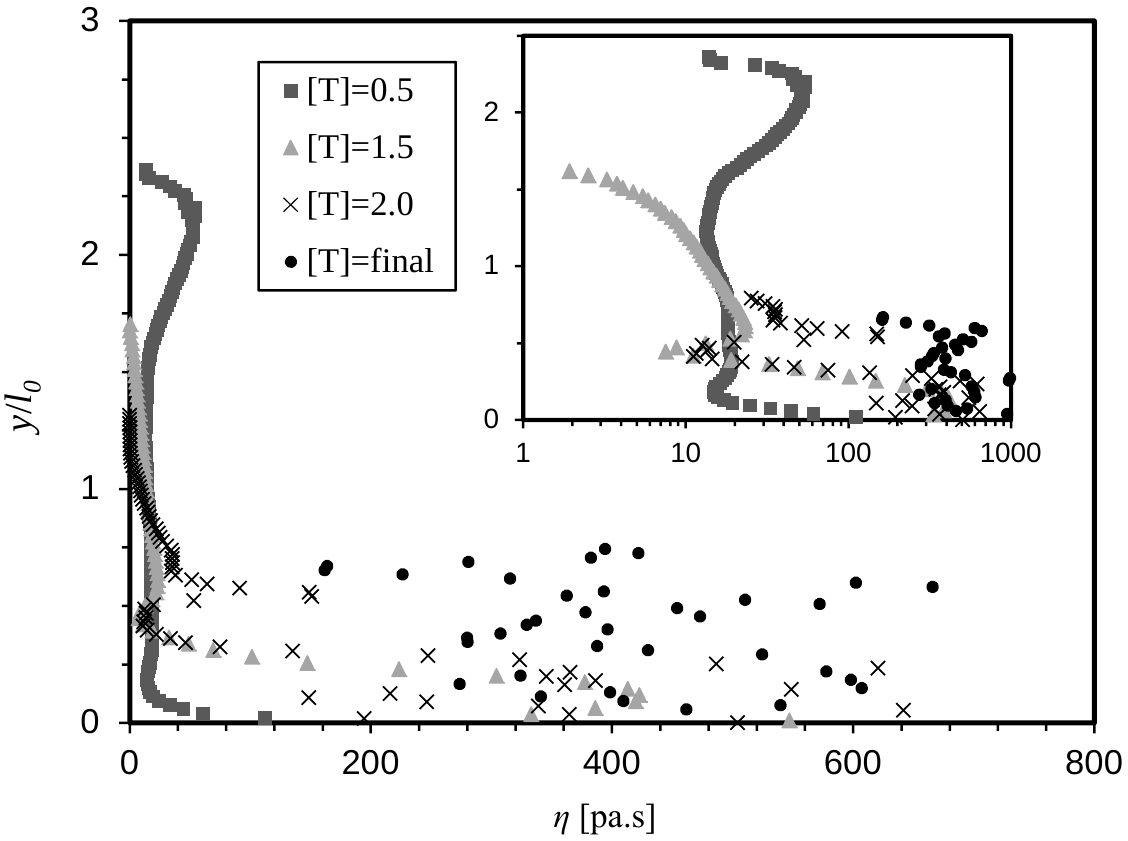}}
\subfloat[][]{\includegraphics[width=6cm]{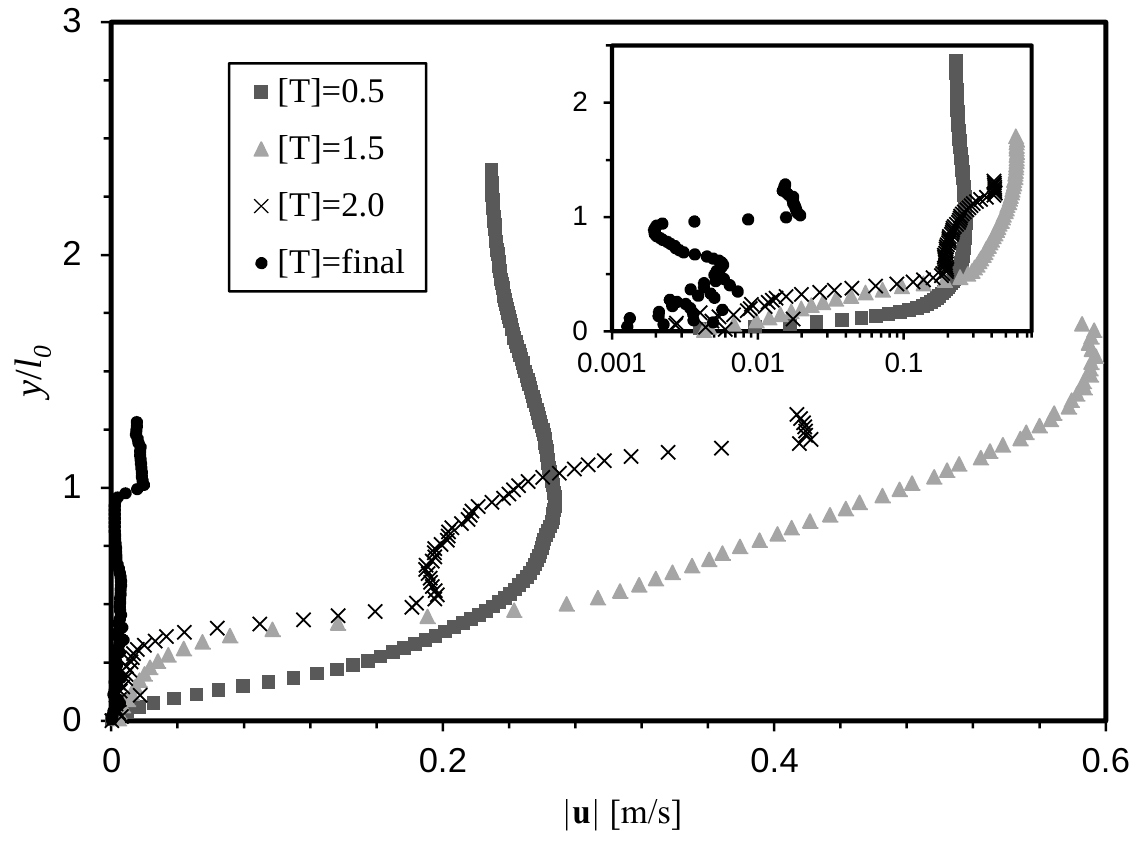}}%
\caption{ Sample profiles of the simulated effective viscosity and velocity magnitude at $x=0.8 l_0$ (inset:logarithimc scale). (a) Effective viscosity with $a=0.6$ (test D1); (b) Velocity magnitude with $a=0.6$ (test D1); (c) Effective viscosity with $a=2.4$ (test D2); (d) Velocity magnitude with $a=2.4$ (test D2).}%
\label{fig:velo_visc}%
\end{figure}

  The collapse of dry granular columns with aspect ratios of $a=0.6$ and $a=2.4$ has been simulated. Figs.\ref{fig:result_all_d1} and \ref{fig:result_all_d2} show the snapshots of the simulation results including the particle configuration, the viscosity field (in logarithmic scale), and the velocity field and the vectors in different dimensionless times ($[T]=t/\surd {h_0/g}$). One can differentiate the yielded (failure) and non-yielded zones by drawing a yield line at locus of the breaking points of vertical layers, or from the velocity and viscous fields (where, the no-yielded zone has a near-zero velocity and high viscosity). For the smaller aspect ratio of $a=0.6$ (Fig.\ref{fig:result_all_d1}), upon release the failure starts by avalanching the frontier face and a trapezoidal non-yielded zone forms at the lower left corner (bounded by the bed, the left wall, a concave yield line and a flat top).  As time goes on, the non-yielded zone grows and approaches the surface until the yielded (failure) zone almost vanishes and the final profile forms.  For larger aspect ratio of $a=0.6$ (Fig. \ref{fig:result_all_d2}), the failure starts by collapsing the upper part of granular assembly (where the vertical acceleration is dominant) with an small triangular not-yielded zone at the lower left corner. The foot of the pile is then propagated along the channel (where horizontal acceleration becomes more dominant) and the not-yielded zone grows and is stretched until it reaches the surface and the final profile forms. One can conclude that the flow dynamic and the final profile depend on the initial aspect ratio granular assembly. To better show the variation of viscosity and velocity, Some example vertical profiles, extracted at $x=0.8l_0$, are provided in Fig. \ref{fig:velo_visc}.\\

\begin{figure}%
\centering
\subfloat[][]{\includegraphics[width=8cm]{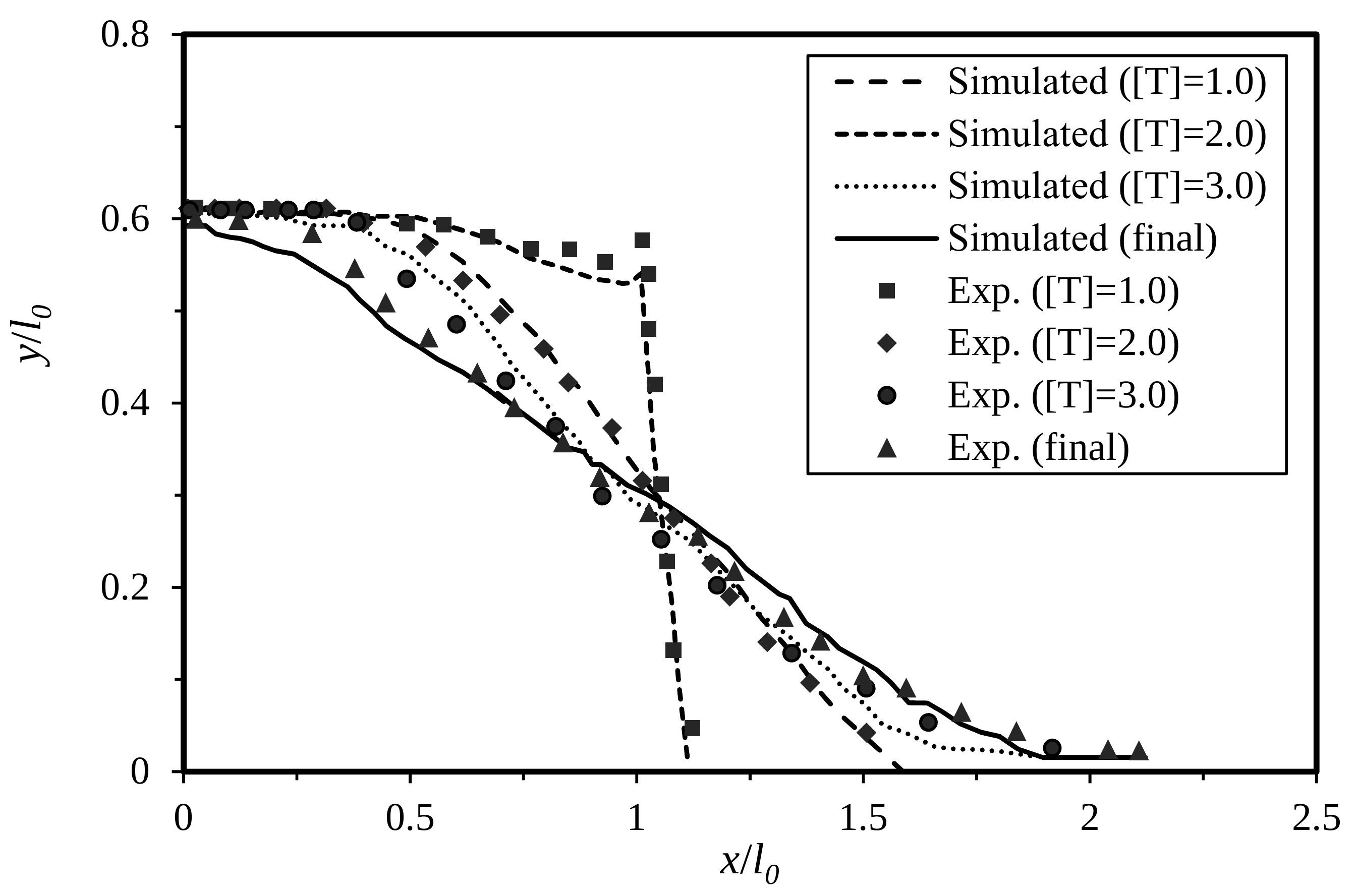}}
\subfloat[][]{\includegraphics[width=8cm]{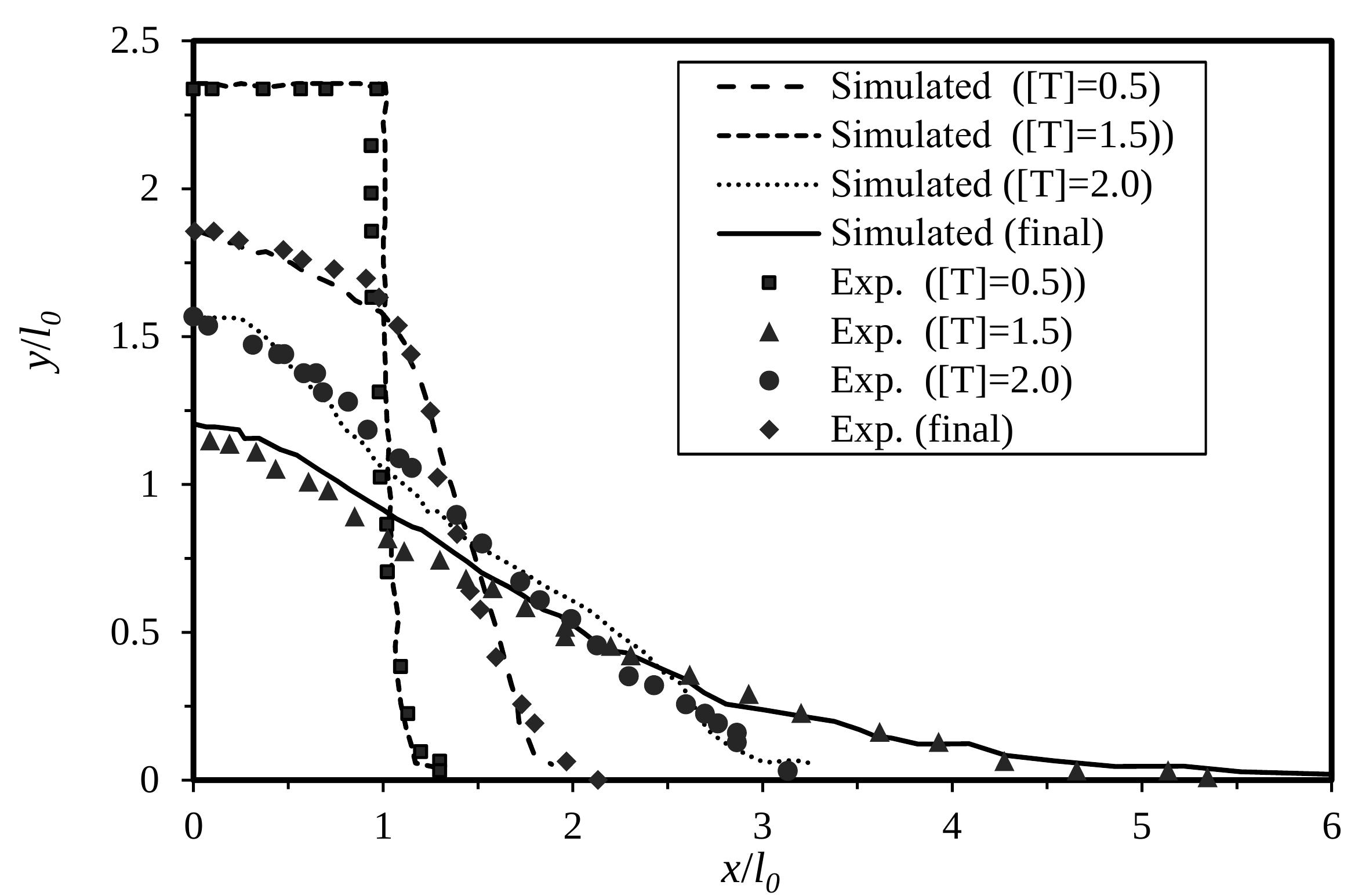}}%

\caption{Numerical and experimental surface profiles for dry granular flow,(a) $a=0.6?$ (b) $a=2.4?$}%
\label{fig:profile_d}%
\end{figure}

\begin{figure}
  \centering
  \includegraphics[width=8cm]{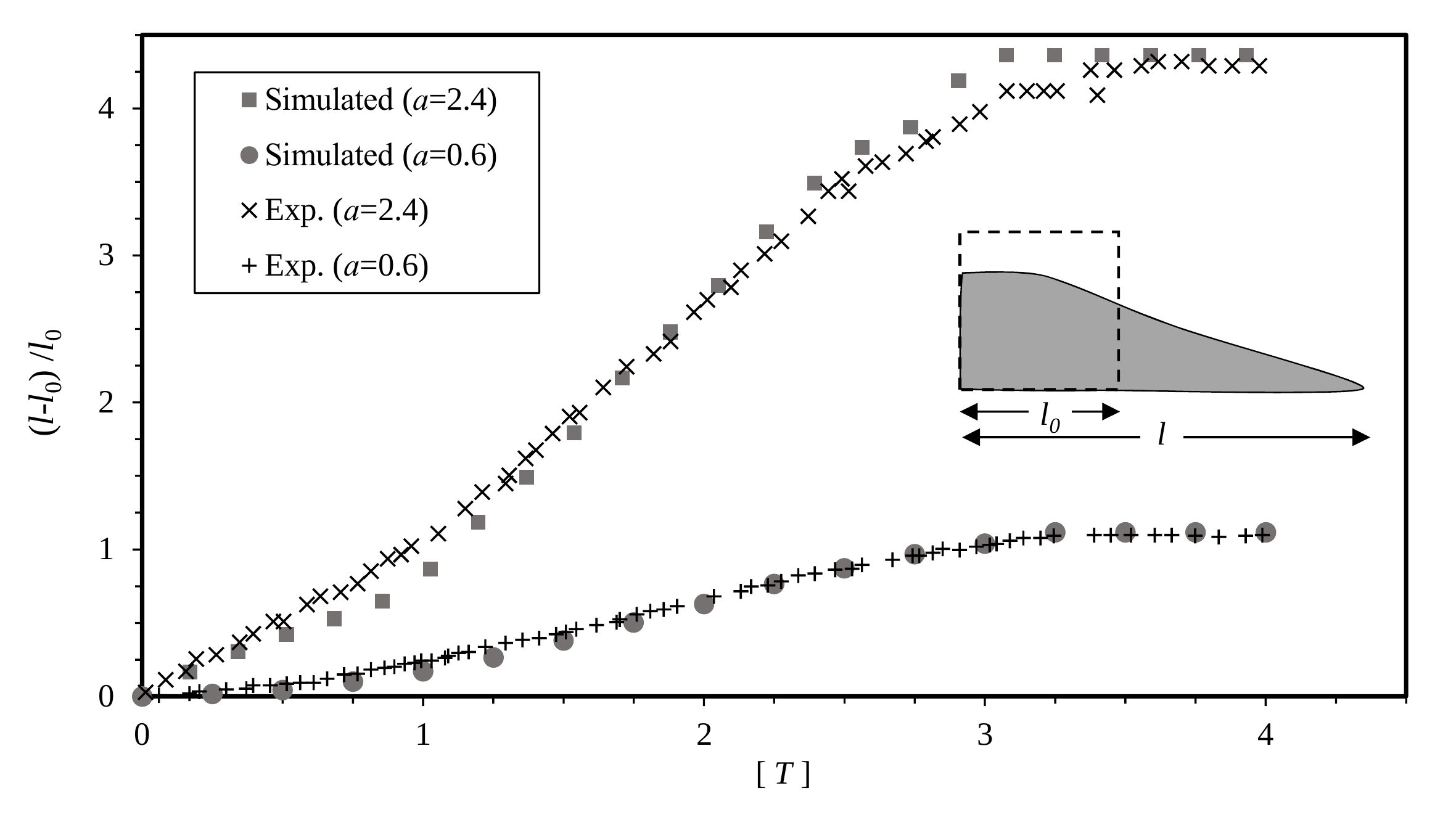}\\
  \caption{Numerical and experimental runout distance of the granular deposit for different $a$ values}
  \label{fig:front_d}%
\end{figure}

To validate the results, the numerical and experimental (digitized from available snapshots in \cite{lajeunesse2005}) surface profiles of the granular assembly are compared in Fig. \ref{fig:profile_d} for four dimensionless times and two aspect ratios. The results show a relatively good agreement between simulations and experiments. The evolution of the granular deposit is also quantified and validated by plotting the dimensionless runout length for different $a$ values (Fig. \ref{fig:front_d}). The position of the flow front in the numerical results is defined by the last particle at the flow toe that is still touching the bulk. Three distinctive regions of (1) accelerating $[T]< \sim1$, (2) steady flow, and (3) decelerating $[T]> \sim3$ regions, can be recognized from this graph. The numerical and experimental values are in a good agreement, although some small discrepancies is observed near deceleration region. Followings, the effects of the various numerical and rheological techniques and parameters are investigated. In each case only the paramenter/technique of interest is examined and the other parameters/techniques are kept as those of the reference model.

\subsubsection{Effect of regularization}
\begin{figure}%
\centering
\subfloat[][]{\includegraphics[width=6cm]{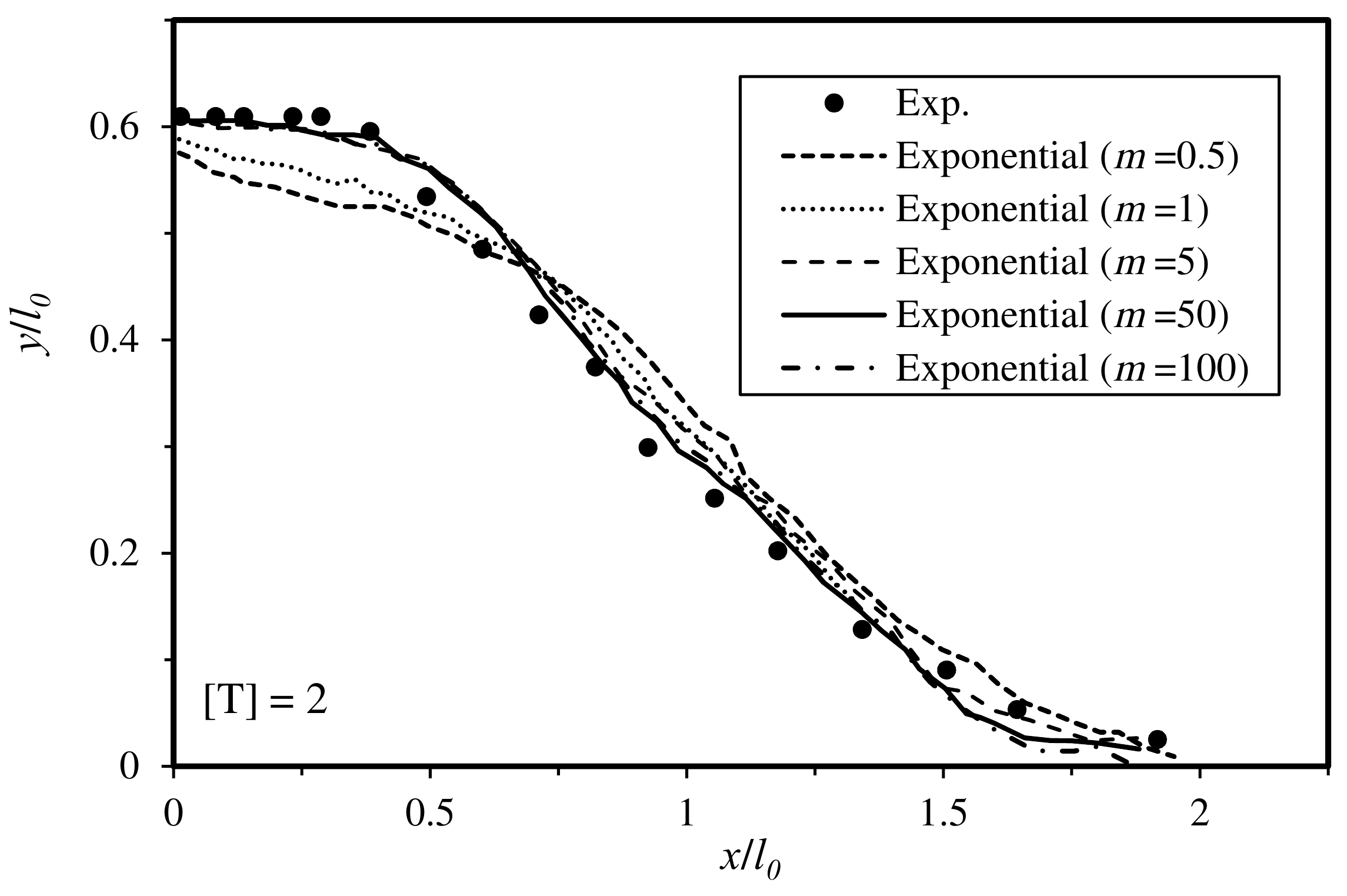}}
\subfloat[][]{\includegraphics[width=6cm]{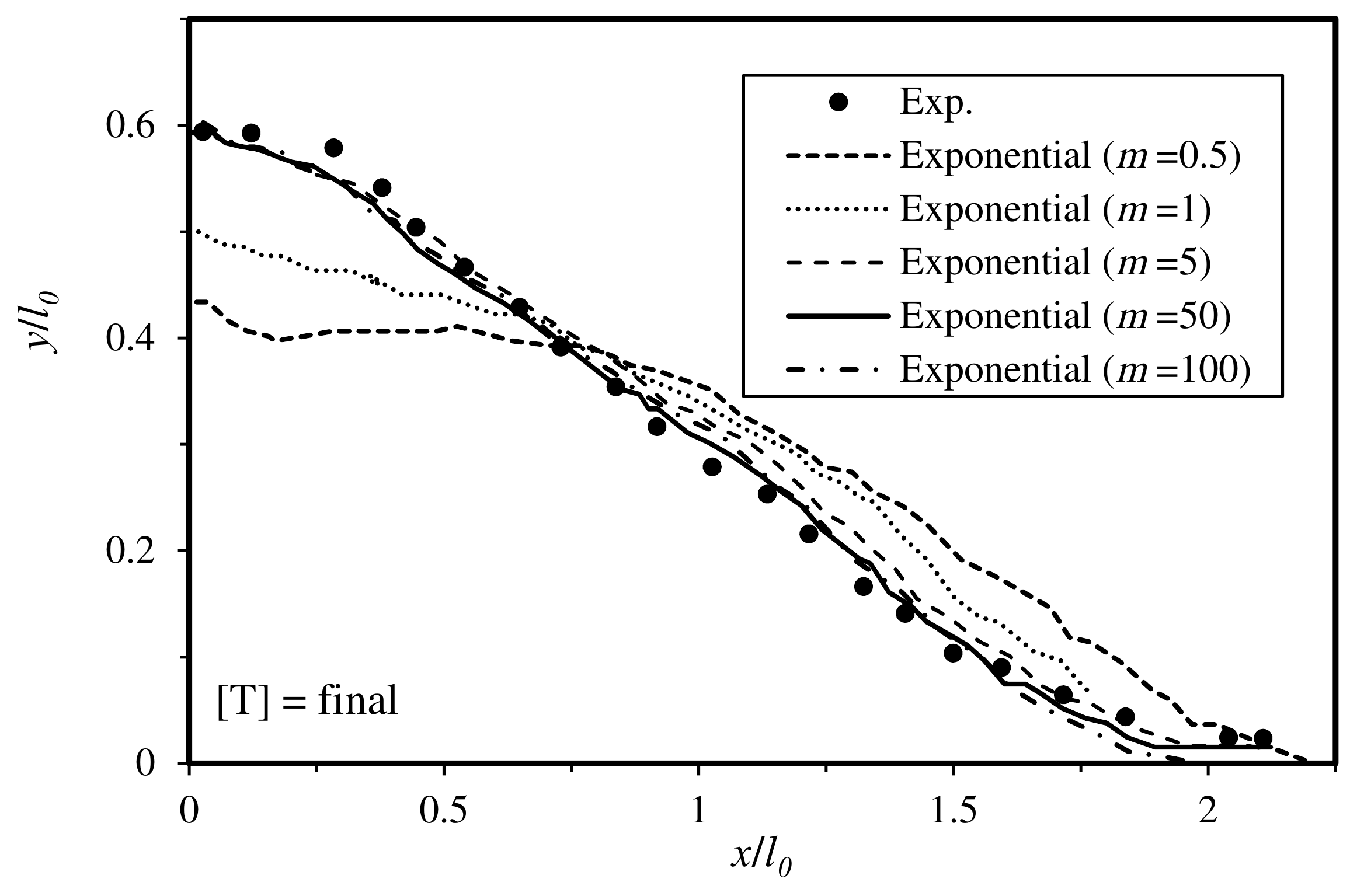}}\\
\subfloat[][]{\includegraphics[width=6cm]{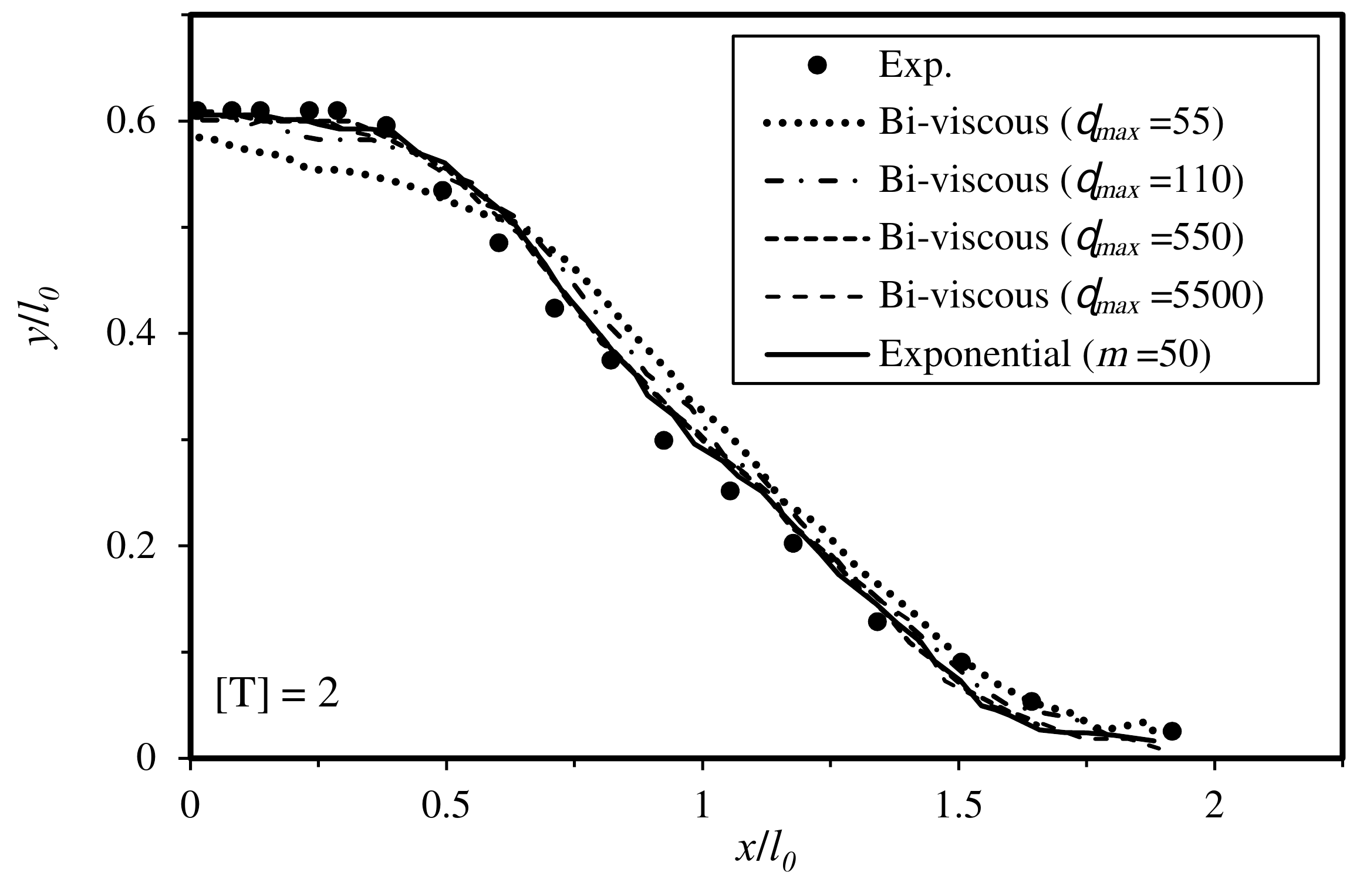}}
\subfloat[][]{\includegraphics[width=6cm]{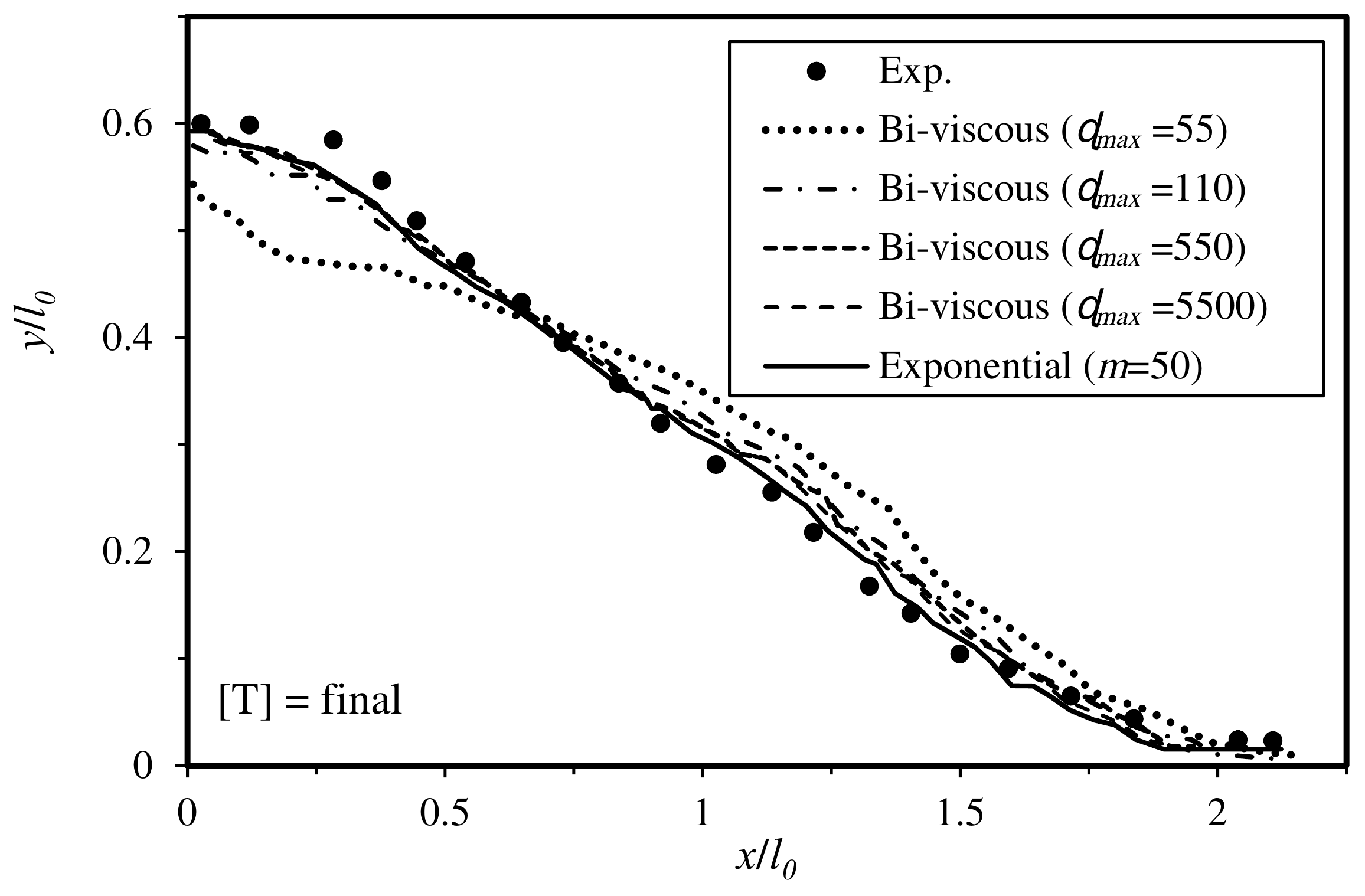}}%
\caption{Effect of the regularization methods and parameters on the surface profile of the dry deposit with $a=0.6$ (test D1) at $[T]=2$ and $[T]=final$}%
\label{fig:mm_graph}%
\end{figure}

To investigate the effect of the regularization methods and parameters, the exponential regularization (Eq. \ref{eq.exponential}) with different exponents, $m$, and bi-viscous regularization (Eq. \ref{eq.bi-viscous}) with different maximum effective viscosities, $\eta_{max}$ are examined. Fig. \ref{fig:mm_graph} (a, b) compares the surface profiles of the granular assembly of case D1 ($a=0.6$) at two example times for five $m$ values. As it is expected, the smaller $m$ values result in a lower maximum viscosity, and a looser collapse. The numerical profiles approach the experimental ones as $m$ increases. The results become independent of $m$ for the $m\geq5$, which corresponds to an average ${\eta _{\max }}\geq\sim 550 pa.s$  (estimated by ${\eta _{\max }} = {\eta _0} + 0.5m{\tau _y}$, ${\tau _y}\approx {\varrho_b gh_0 sin(\theta_r)/2}$).\\

Fig. \ref{fig:mm_graph} (c, d) compares the surface profiles resulted from the exponential regularization (with $m=50$) and bi-viscous regularization with $\eta_{max}\sim$ 55, 110, 550 and 5500 (equivalent to $m=$ 0.5, 1.0, 5.0 and 50, respectively). As the figure shows, by increasing the $\eta_{max}$ values the results of the bi-viscous regularization approach those of experimental. The bi-viscous regularization(with $\eta_{max}\approx 5500$), produces profiles similar to the experiments. Comparisons show that the selection of the appropriate values for exponent $m$ or maximum effective viscosities, $\eta_{max}$, is more important than the choice regularization method for reproducing the true failure mechanism.

%

\subsubsection{Effect of the rheological parameters}

\begin{figure}%
\centering
\subfloat[][]{\includegraphics[width=6cm]{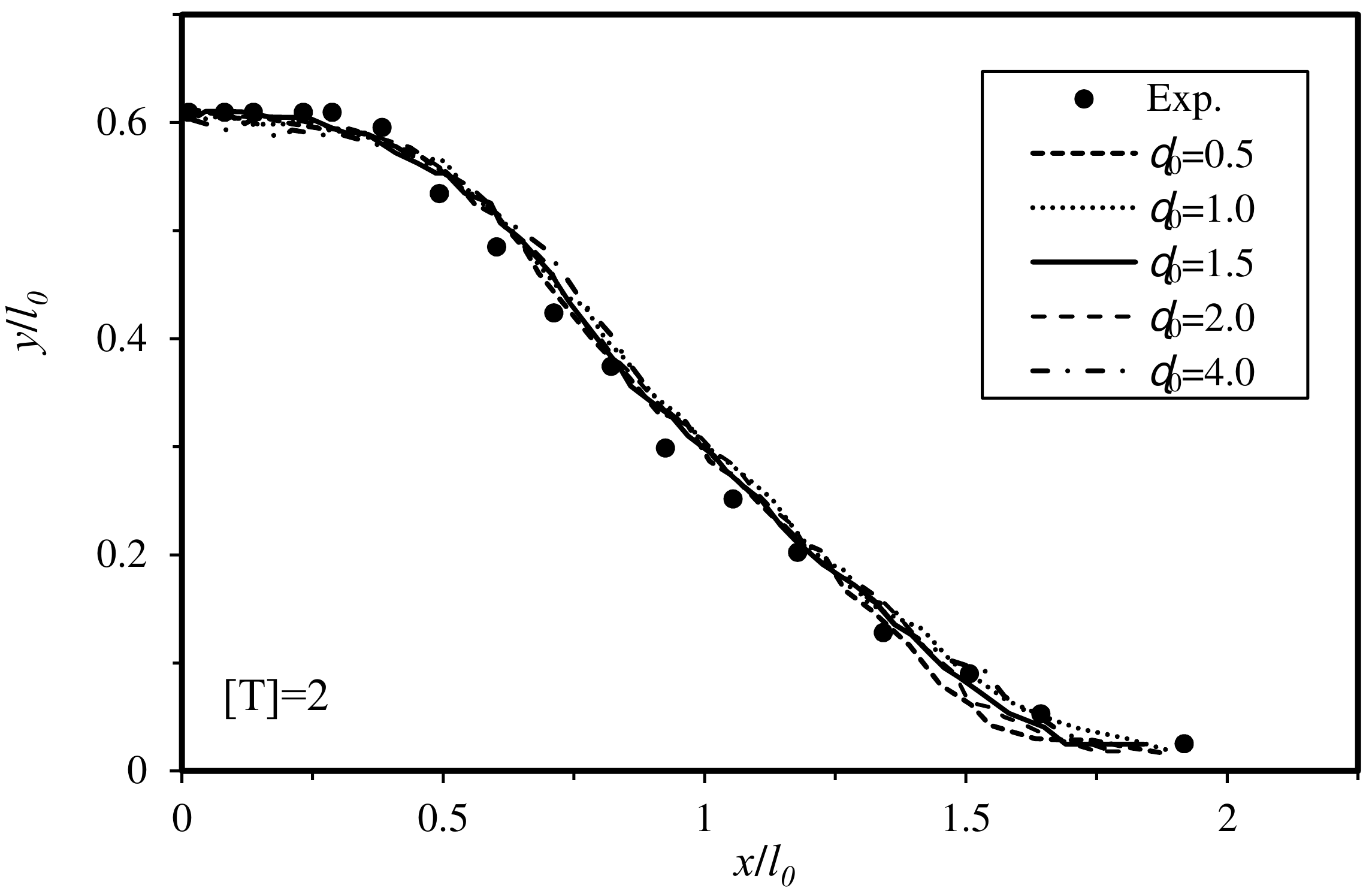}}
\subfloat[][]{\includegraphics[width=6cm]{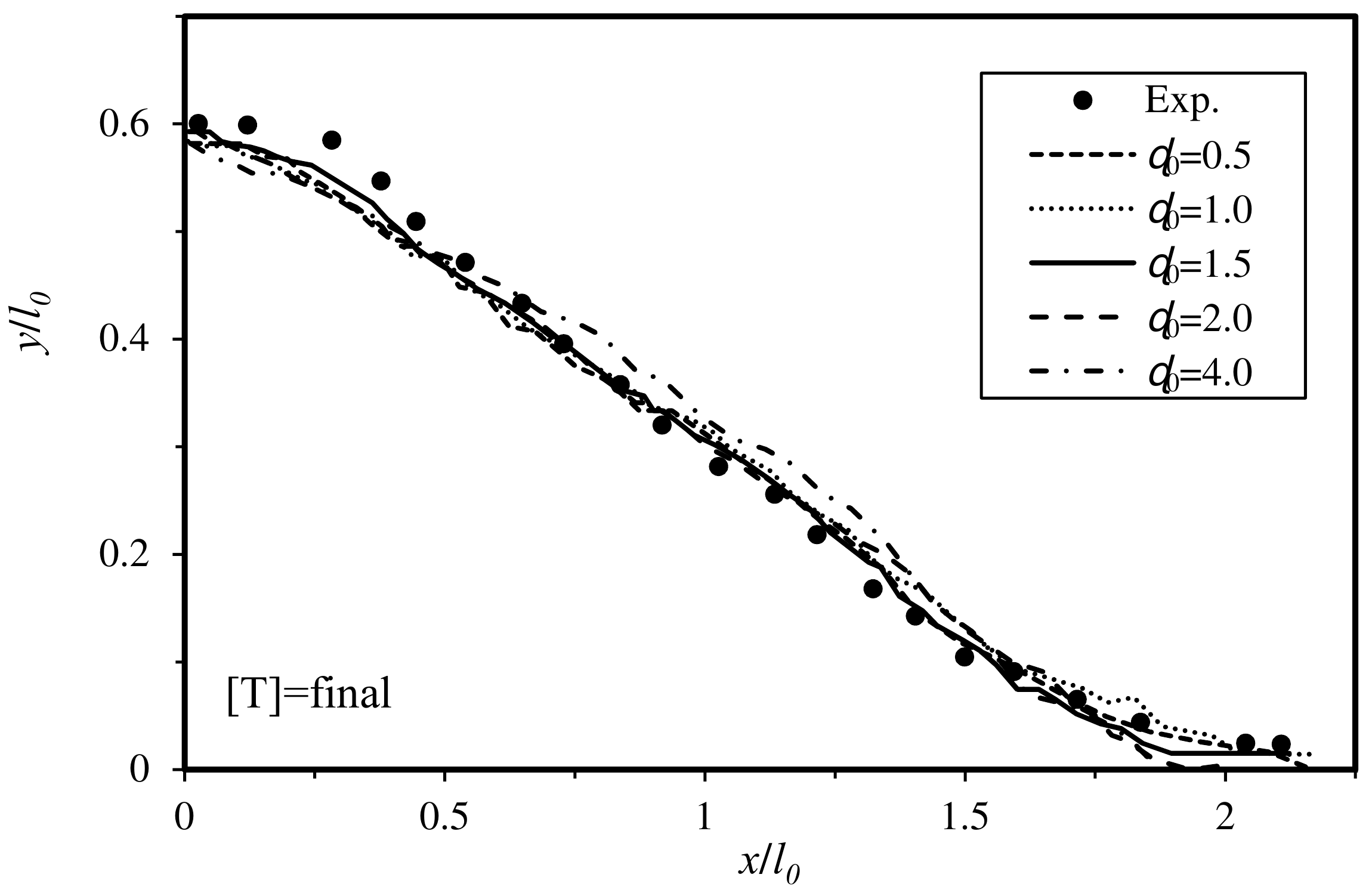}}\\
\subfloat[][]{\includegraphics[width=6cm]{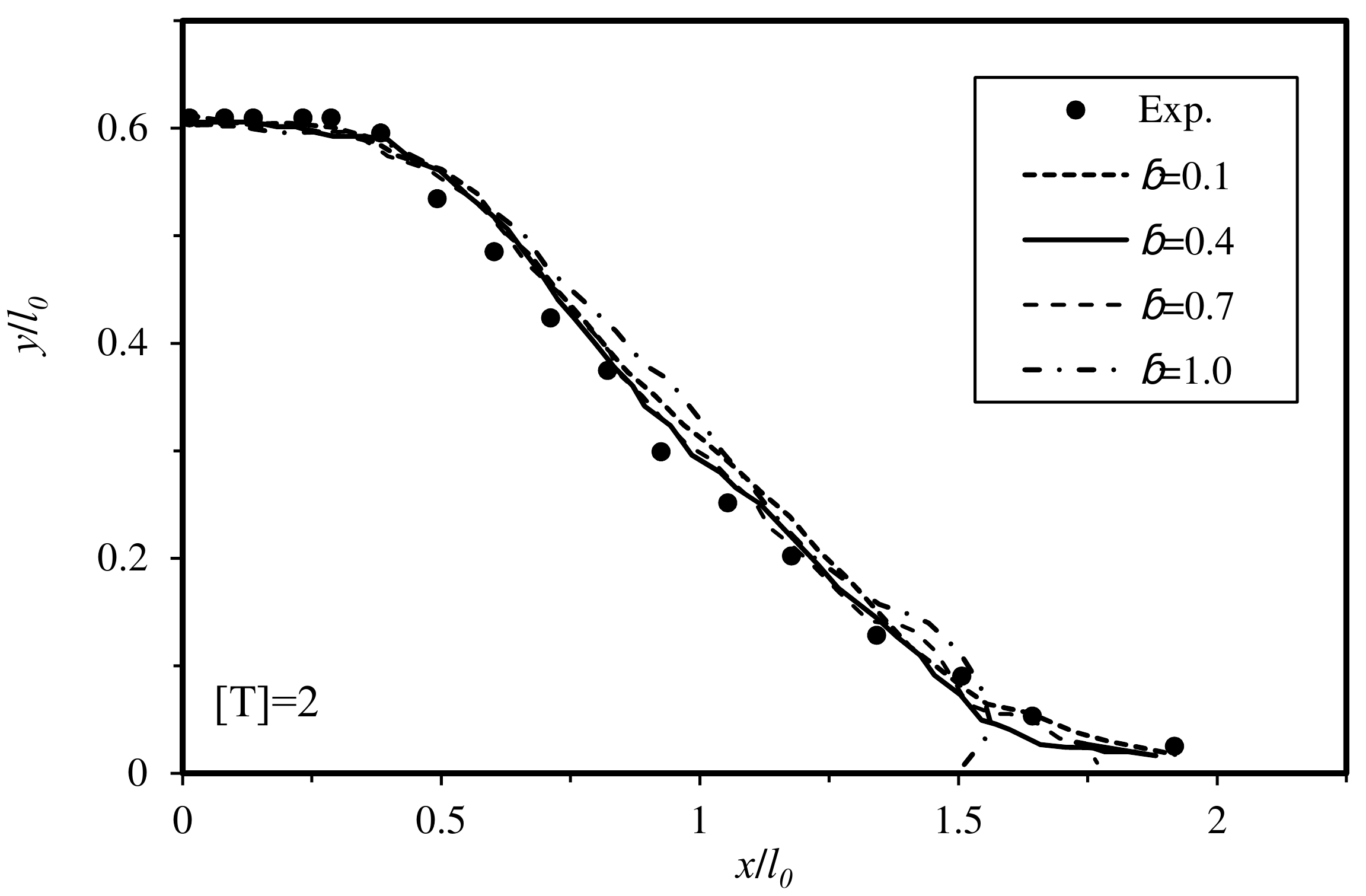}}
\subfloat[][]{\includegraphics[width=6cm]{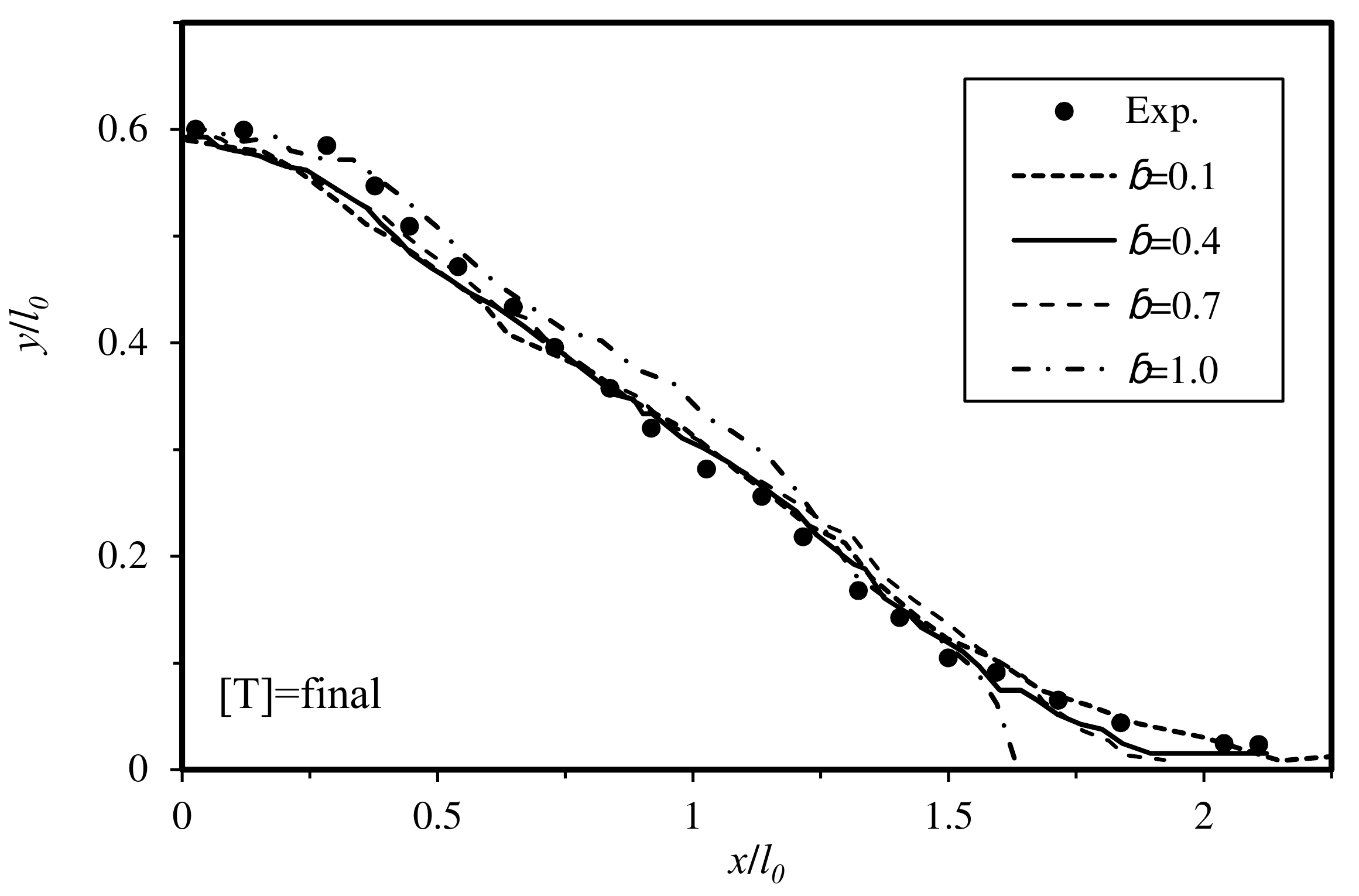}}
\caption{Effect of the flow behaviour index $\beta$ and flow consistency index $\eta_0$ on surface profile of the dry deposit with $a=0.6$ (case D1) at $[T]=2$ and $[T]=final$}%
\label{fig:beta_etha_graph}%
\end{figure}

In absence of the rheometry measurements, it is important to analyse the sensitivity of the numerical results to the choice of rheological parameters. The effect of the post-failure rheological parameters, i.e. the consistency index $\eta_0$ and flow behavior index, $\beta$, are investigated here. Fig. \ref{fig:beta_etha_graph} (a, b) compares the surface profiles for different $\eta_0$ values (ranging from 0.5 to 4.0 $pa.s^\beta$). As the figure shows, the choice of flow consistency index $\eta_0$ has an insignificant effect on the results. This can be due to the fact that $\eta_0$ value is a very small portion of the overall effective viscosity (see \ref{fig:velo_visc} a and c).
Fig. \ref{fig:beta_etha_graph} (c, d) compares the surface profile of granular assembly for the models with $\beta$ values (accounting for the nonlinearity) ranging from 0.1 to 1.0. As the results show the choice of $\beta$ affects the surface profiles (especially the finals). Although $\beta=1.0$ predict an accurate surface profile at the top, it underestimates the runout length. $\beta=0.4$ and $0.7$ produces the similar (and more accurate) surface profiles and runout lengths, compatible with the experiments (although the top flatness is slightly less than experiments). $\beta=0.1$  produce a loose triangular (not truncated) profile with an overestimated runout length and uneven top.

%

\subsubsection{Effect of the normal stress}

\begin{figure}%
\centering
\subfloat[][]{\includegraphics[width=6cm]{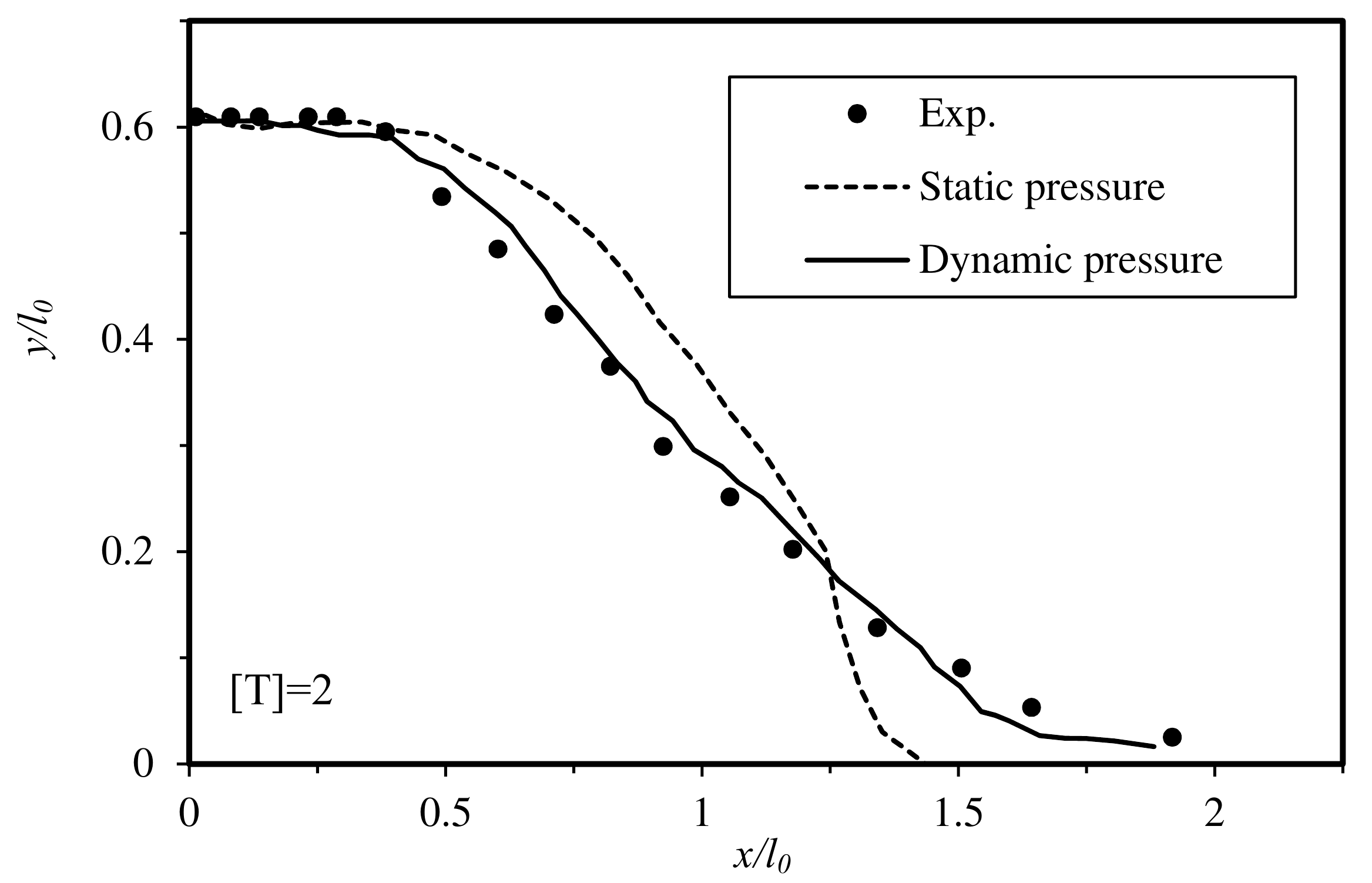}}
\subfloat[][]{\includegraphics[width=6cm]{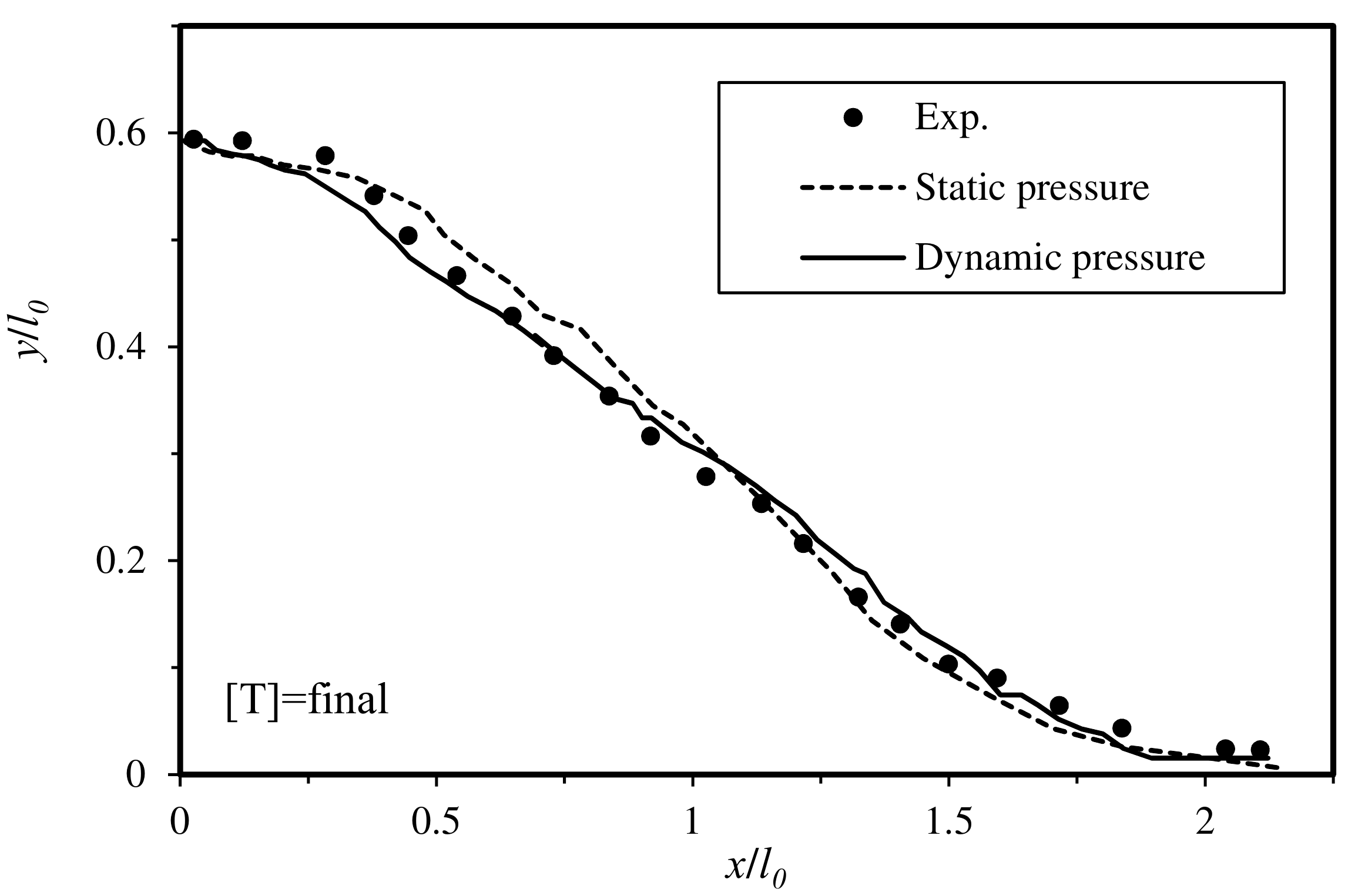}}%

\caption{Effect of static and dynamic effective pressure $p'$ on surface profile of the dry deposit with $a=0.6$ (test D1) at: (a) $[T]=2$, (b) $[T]=final$.}%
\label{fig:effective_graph}%
\end{figure}

Fig. \ref{fig:effective_graph} shows the effect of the normal stress (effective pressure) calculation method (for case D1). An effective pressure based on the proposed smoothed dynamic pressure successfully reproduces the experimental profiles. The commonly used static pressure has a larger avalanche angle, especially at the initial stages. That is due to fact that an static pressure ignores the vertical acceleration (which is important in earlier stages of the collapse), therefore overestimates the effective pressure and yield stress.

\subsubsection{Effect of the shear stress calculation method}
\begin{figure}%
\centering
\subfloat[][]{\includegraphics[width=6cm]{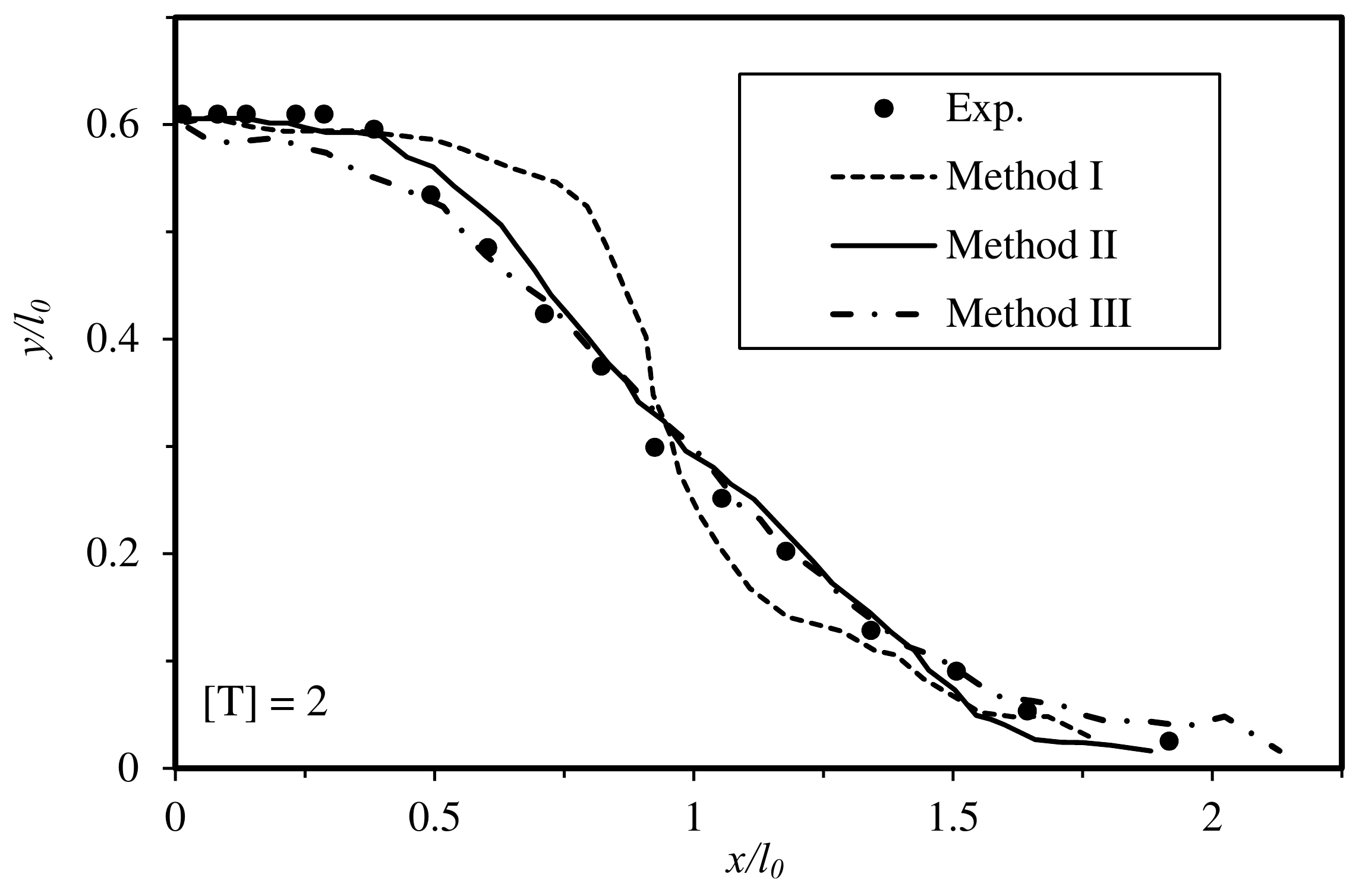}}
\subfloat[][]{\includegraphics[width=6cm]{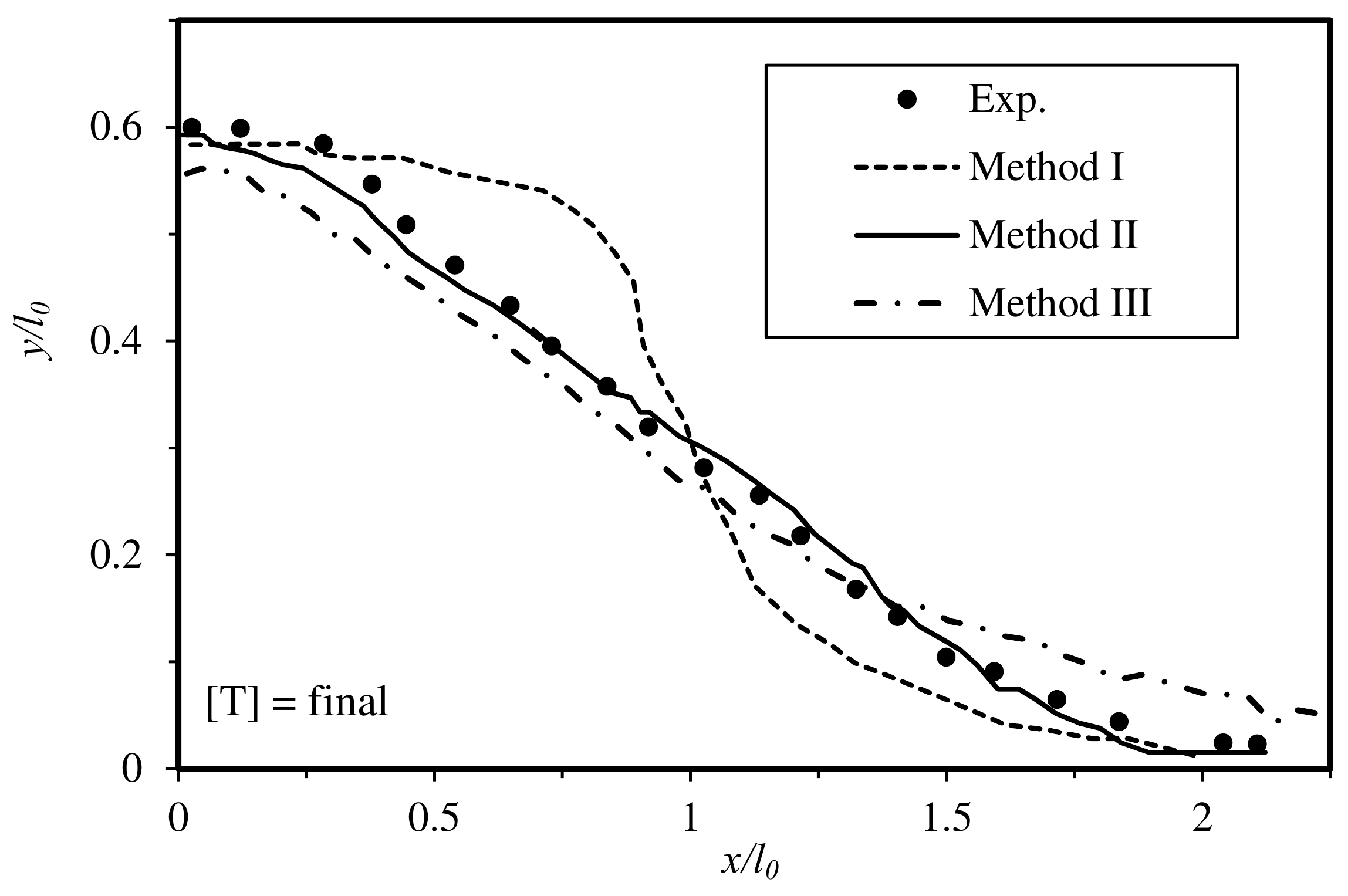}}%

\caption{Effect of the interaction viscosity $\mathord{\buildrel{\lower3pt\hbox{$\scriptscriptstyle\frown$}}\over \eta } _{ij}$ calculation method on surface profile of dry deposit with $a=0.6$ (test D1) at: (a) $[T]=2$, (b) $[T]=final$.  }%
\label{fig:interaction_graph}%
\end{figure}

\begin{figure}
  \centering
  \includegraphics[width=8cm]{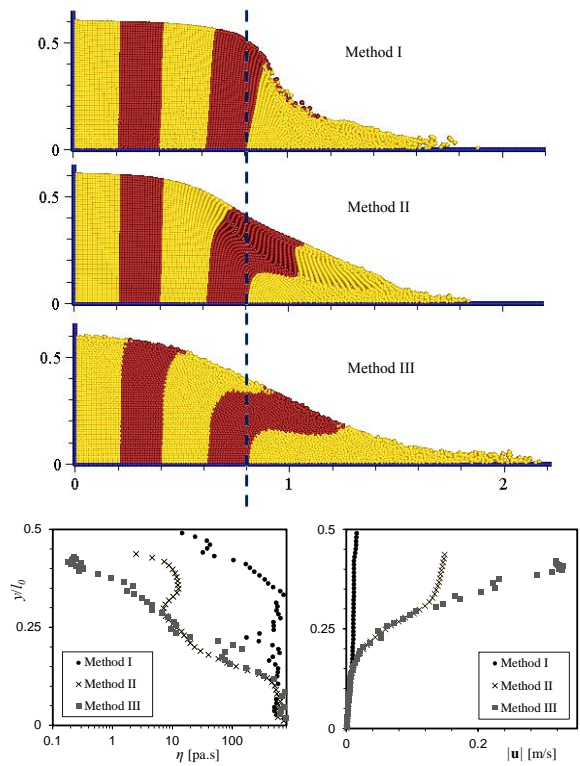}\\
  \caption{Snapshots of the simulated granular deposits at [T]=3 and viscosity and velocity profiles (at $x/l_0=0.8$) for the three methods of the shear stress divergence approximation.}\label{fig:interaction_plot}
  \label{fig:scale_graph}%
\end{figure}

Here the effect of three different methods for the MPS approximation of the divergence of shear stress tensor are investigated. As it was discussed in section \ref{sec.shear_stress},  method I (Eq. \ref{eq.SS1})  directly uses velocity field and the effective viscosity of the target particles to approximate the divergence of shear stress tensor. Method II (Eq. \ref{eq.SS2}) is similar to method I, but it uses a harmonic mean for interaction viscosity ${{\mathord{\buildrel{\lower3pt\hbox{$\scriptscriptstyle\frown$}}\over \eta } }_{ij}} =  {{2{\eta _i}{\eta _j}} \mathord{\left/
 {\vphantom {{2{\eta _i}{\eta _j}} {\left( {{\eta _i} + {\eta _j}} \right)}}} \right.\kern-\nulldelimiterspace} {\left( {{\eta _i} + {\eta _j}} \right)}}$.  Method III (Eq. \ref{eq.SS3}) still uses an harmonic mean for the interaction viscosity but it first approximates the shear stress tenors then calculates its divergence. The effect of these three methods on the surface profile of the granular assembly (for case D1) is shown in Fig. \ref{fig:interaction_graph}. Fig. \ref{fig:interaction_plot} also compares  sample snapshots and viscosity/velocity profiles for these methods. Results show that the method I overestimates the near-surface viscosity and creats a steeper avalanche angle. Among these three methods the Method II produces the most accurate surface profiles. Method III,  smoothes the abrupt variations in the viscosity and velocity fields resulting in a slightly looser granular deposit. This can be because, the MPS integration (kernel smoothing) has been applied twice in this method to approximate the divergence of the shear stress tensor.

\subsubsection{Scalability}
\begin{figure}%
\centering
\subfloat[][]{\includegraphics[width=6cm]{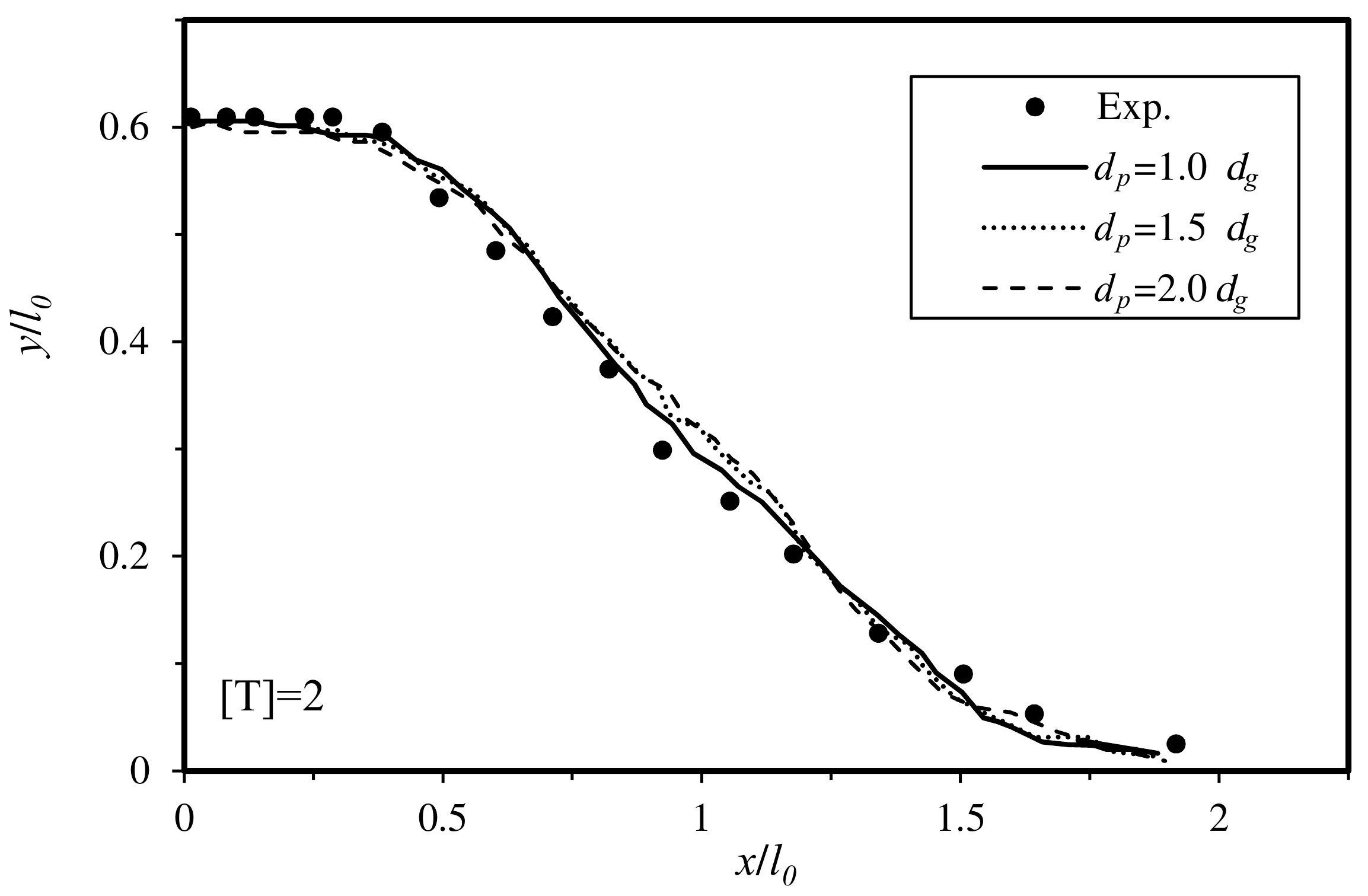}}
\subfloat[][]{\includegraphics[width=6cm]{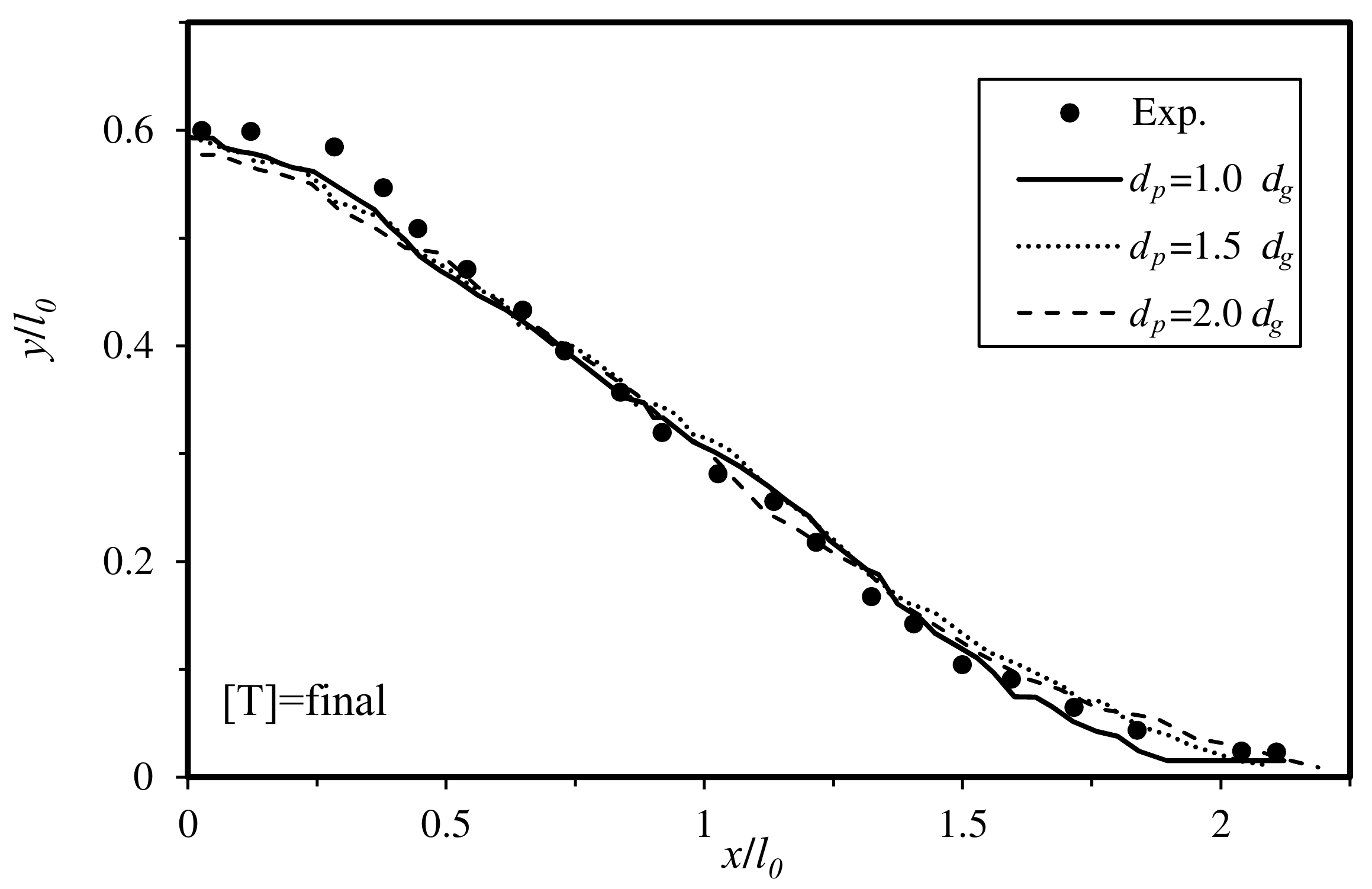}}%

\caption{Effect of the particle size $d_p$ on the surface profile of the dry deposit with $a=0.6$ (test D1) at: (a) $[T]=2$, (b) $[T]=final$.  }%
\label{fig:scale_snapshot}%
\end{figure}

An important advantage of continuum models, comparing to discrete models like DEM, is their scalability (computationally affordable for larger scale problems). To evaluate the scalability of the proposed continuum-based model, various particle sizes ($d_p=$ 1.0, 1.5 and 2.0 times of the grain size $d_g$) are examined. Figs. \ref{fig:scale_graph} and \ref{fig:scale_snapshot} compare the surface profiles and example snapshots, achieved from different particle sizes (for case D1), respectively. As the figures show, using particle sizes larger than the grain size does not have a significant effect on the simulation results, yet decreases the computational cost, as the number of particles decreases. Therefore, the model is applicable to larger scale problem using the computationally affordable number of particles, regardless of the grain sizes.

\begin{figure}
  \centering
  \includegraphics[width=8cm]{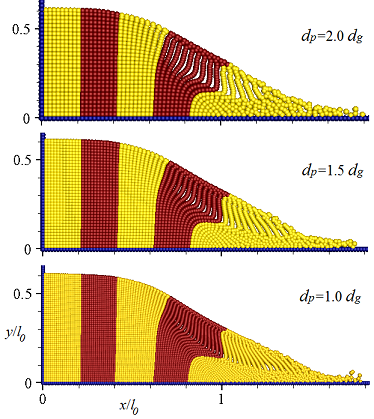}\\
  \caption{Configuration of the granular deposits at [T]=2.5 for three particle sizes}\label{fig:results1}
  \label{fig:scale_graph}%
\end{figure}

\subsection{Submerged granular flow results}

\begin{figure}
  \centering
  \includegraphics[width=\textwidth]{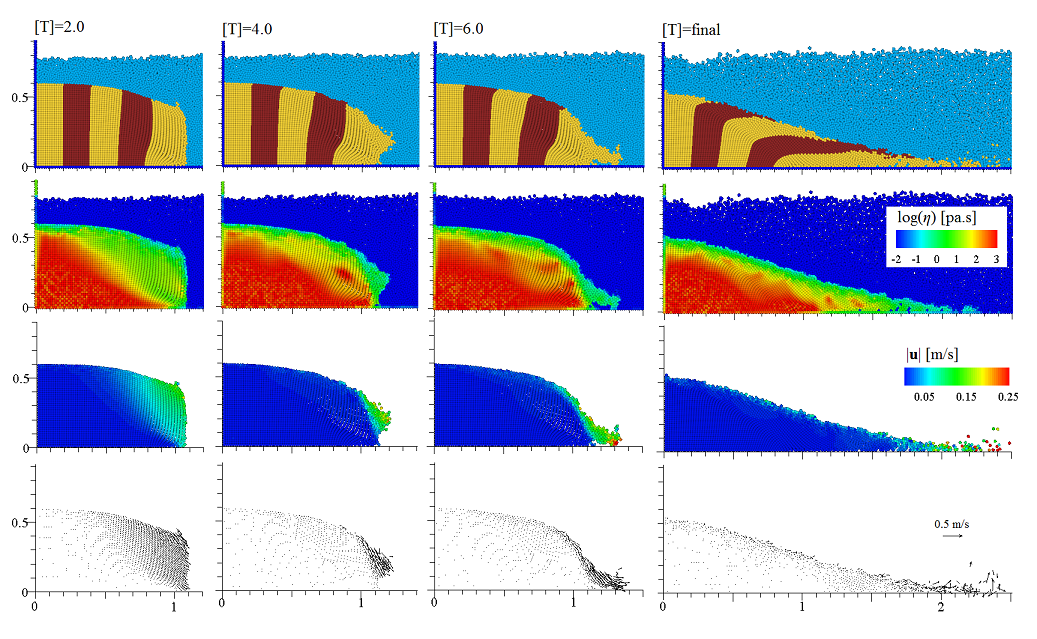}\\
  \caption{Deposit configuration, viscous field and velocity field and vectors for submerged granular collapse with $a=0.6$  (test S1)}\label{fig:results_configd1}
  \label{fig:result_all_s1}%
\end{figure}

\begin{figure}
  \centering
  \includegraphics[width=\textwidth]{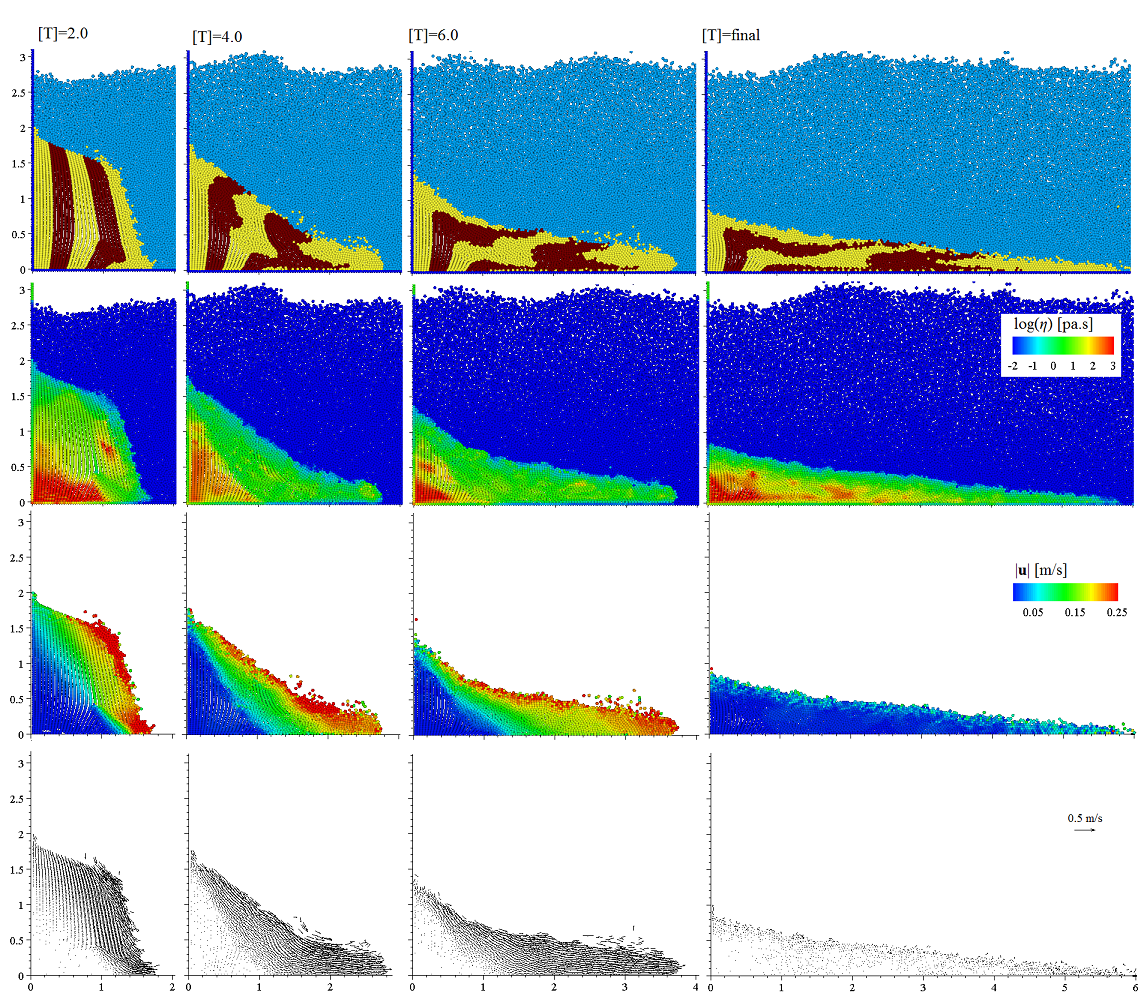}\\
  \caption{Deposit configuration, viscous field and velocity field and vectors for submerged granular collapse with $a=2.4$  (test S2)}\label{fig:resultsd_configd2}
  \label{fig:result_all_s2}%
\end{figure}

The developed and calibrated model of this study is also applied to the collapse of submerged granular columns with aspect ratios of $a=0.6$ and $a=2.4$. A particle size $d_p=2d_g$ has been used. Considering the geometrical and material properties (Tables \ref{table:geo} and \ref{table:material}) both cases are within the inertia regime of the granular flow. Figs. \ref{fig:result_all_s1} and \ref{fig:result_all_s2} illustrate the time evolution of the deposit configuration, the viscous field (in logarithmic scale), the velocity field (for granular phase), and the velocity vectors for  $a=0.6$ and $a=2.4$, respectively. Similar to the dry case, the results are provided for different dimensionless times $[T]=t/\surd {h_0/g}$. The overall configuration and failure mechanism of the submerged granular deposit is similar to that of the dry case, but with a longer time scale (about 2.5 times), smaller velocity, thicker front and more fluctuating granular surface. For the smaller aspect ratio of $a=0.6$, a trapezoidal shape deposited with a flat top (slightly wider than that of the dry case) is observed. For larger aspect ratio of $a=2.4$, the collapse results in a triangular-shape deposit with a concave top, similar to the dry case.\\

\begin{figure}
  \centering
  \includegraphics[width=8cm]{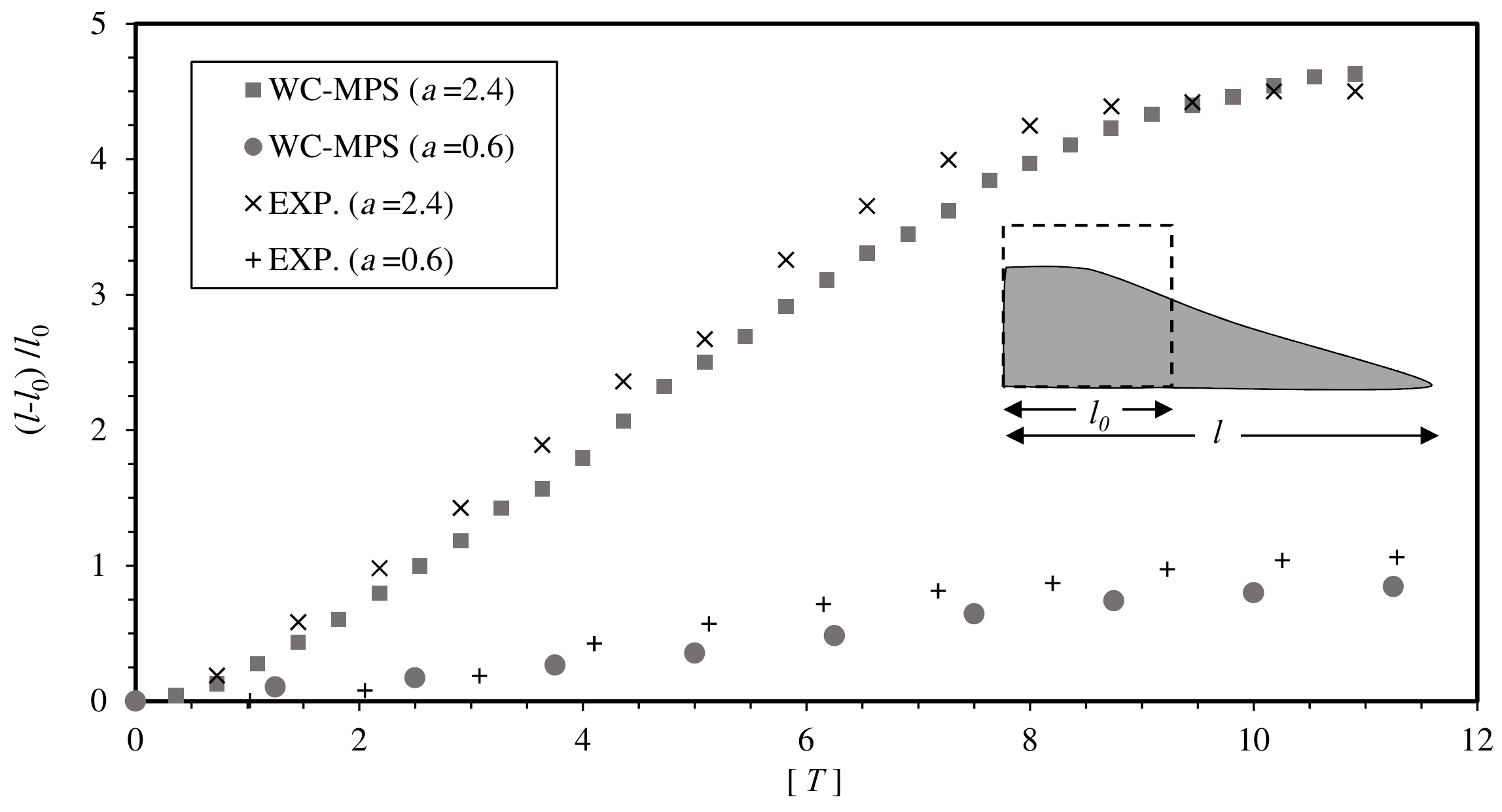}\\
  \caption{Numerical and experimental runout distance of the submerged granular deposits}
  \label{fig:front_s}%
\end{figure}

To validate the results, the numerical and experimental surface profiles of the granular assembly are compared in Fig. \ref{fig:profile_d}, for four dimensionless times and two aspect rations, showing relatively good agreement. However, the flat top of the deposits (for the case with $a=0.6$) in the simulations is narrower than that of the experiments. The discrepancy can be due to some unphysical fluctuations in the fluid phase that erodes down the flat top of the deposit. Also some discrepancies in the earlier time step (for both cases), can be due to the gate removal effect in the experiments, where the gate friction suspends the nearby grains in the ambient fluid. The evolution of the granular deposit shape is also quantified and validated by plotting the numerical and experimental dimensionless runout length for different aspect ratios (Fig. \ref{fig:front_s}). Similar to the dry cases, three distinctive regions (but with a longer time scale) including (1) accelerating $[T]< \sim2$, (2) steady flow, (3) decelerating $[T]> \sim8$ regions, can be recognized. The numerical and experimental values are in a good agreement, although the numerical model results slightly underestimate the experimental values.

\begin{figure}%
\centering
\subfloat[][]{\includegraphics[width=8cm]{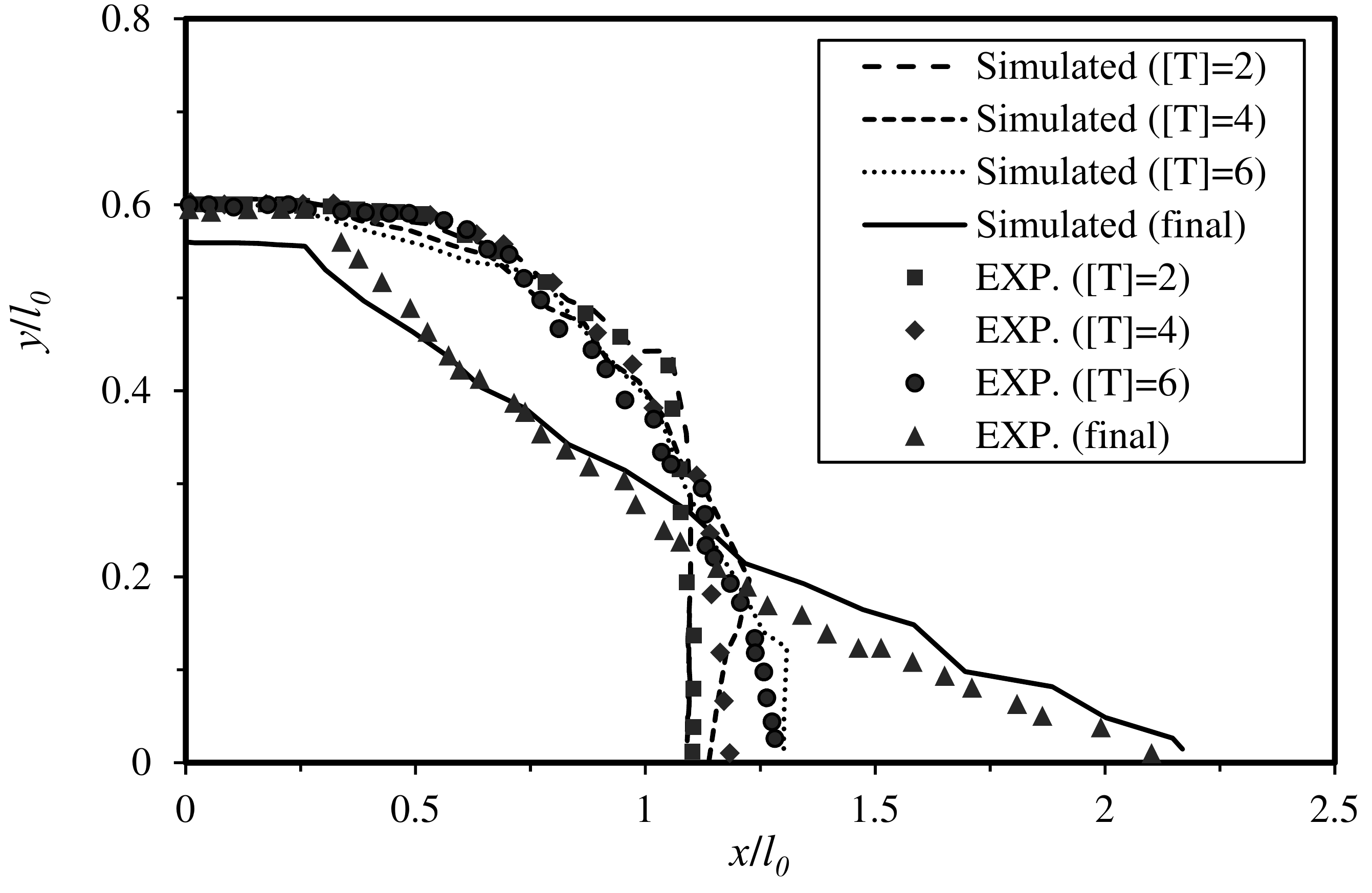}}
\subfloat[][]{\includegraphics[width=8cm]{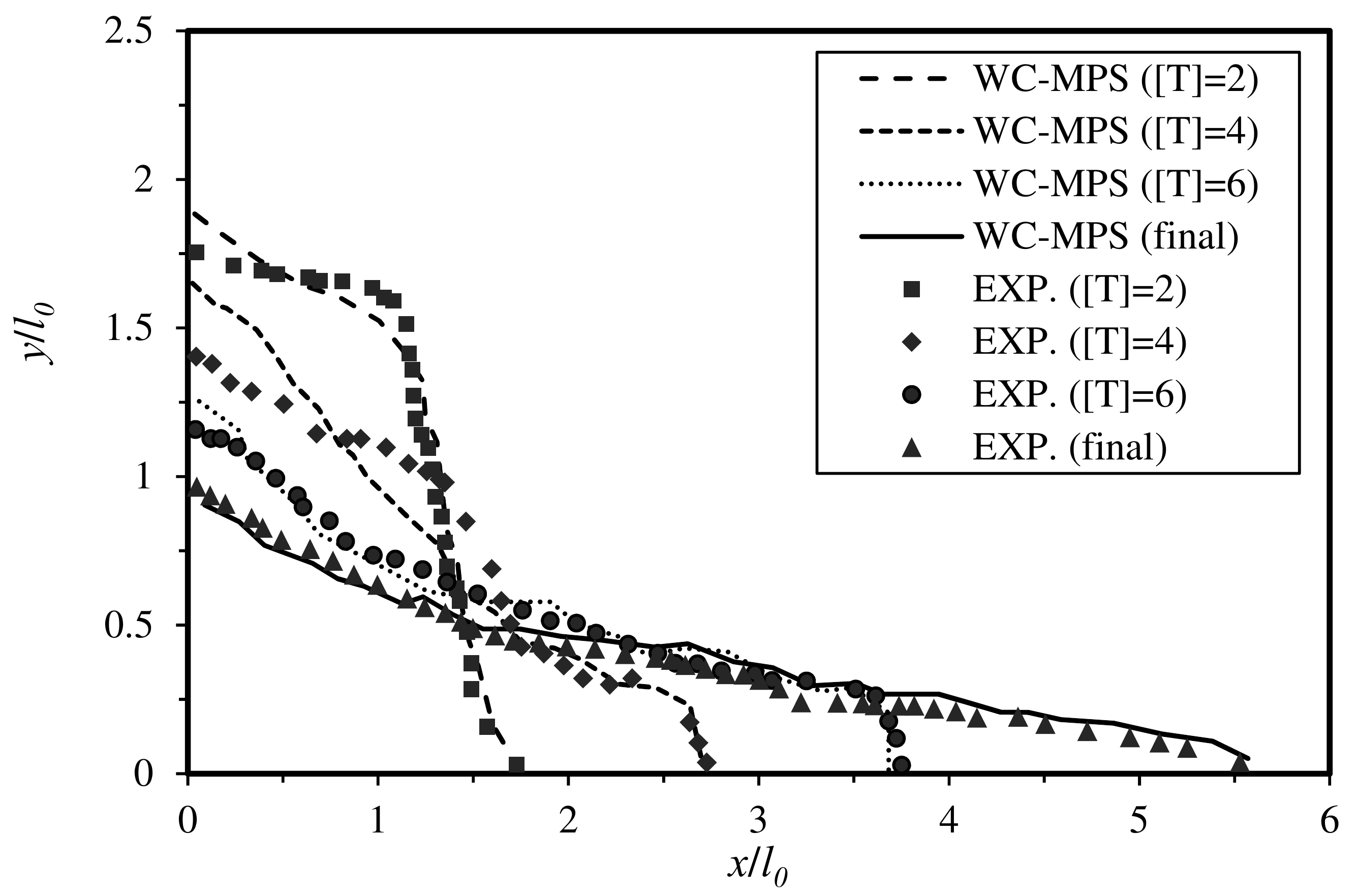}}%

\caption{Numerical and experimental surface profiles of submerged granular flow, (a) $a=0.6$ (b) $a=2.4$}%
\label{fig:profile_s}%
\end{figure}

\section{Conclusion}
A WC--MPS meshfree Lagrangian model for continuum-based numerical modeling of dry and submerged granular flows was developed and fully characterized. The model treated the multiphase system of granular material and the ambient fluid as as a multi-density multi-viscosity system. The viscous behaviour of the granular phase was predicted using a regularized Herschel–-Bulkley rheological model with the Drucker--Prager yield criterion. The numerical algorithms for calculation of the effective viscosity, the effective pressure, and the shear stress divergence were introduced and evaluated.

Validation of the model for the case of viscoplastic Poiseuille flow, showed the capabilities of the model in reproducing the analytical viscoplastic and yield behaviours. The model is then validated and fully characterized (in respect to the various rheological and numerical parameters) for dry and submerged granular collapses with different aspect ratios. The results showed the ability of the models to accurately simulate the granular flow features for both dry and submerged cases. The effective pressure and shear stress calculation methods, as well as regularization parameters, were found to have a significant impact on the accuracy of results. However, the effect of post-failure rheological parameters and particle size on the results were found to be insignificant. Comparison of dry and submerged granular flow revealed the importance role of the ambient fluid in the shape and evolution of the granular deposites and the time scale of collapse. 

\section{Acknowledgement}

Authors would like to thank Dr. O. Pouliquen and Dr. P. Aussillous for providing the video data of their submerged granular collapse experiments.





\bibliographystyle{model6-num-names}
\bibliography{mybib}











\end{document}